\documentclass{aa}
\usepackage[varg]{txfonts}
\usepackage{makecell}
\usepackage{multirow}
\usepackage{natbib}
\usepackage[english]{babel} % remove datetime2 complaints about lang not being defined
\usepackage[utf8]{inputenc}
\usepackage[T1]{fontenc}
\usepackage{xspace}
\xspaceaddexceptions{”}
\usepackage[calc]{datetime2}
\usepackage{hyperref}
\usepackage{xcolor}

\usepackage{savesym}
\savesymbol{tablenum}
\usepackage{siunitx}
\restoresymbol{SIX}{tablenum}
\DeclareSIUnit\day{\text{day}}
\DeclareSIUnit\minute{\text{minute}}
\DeclareSIUnit\pixel{\text{pixel}}

\hypersetup{%
  colorlinks=true,
  linkcolor=blue,
  urlcolor=blue,
  citecolor=blue,
  bookmarksnumbered=true,
  bookmarksopenlevel=1,
  }
\usepackage{etoolbox}
\makeatletter
\patchcmd\@combinedblfloats{\box\@outputbox}{\unvbox\@outputbox}{}{}
\makeatother

\newcommand\ie{i.e.\xspace}
\newcommand\eg{e.g.\xspace}
\newcommand\kmps{\kilo\meter\per\second}
\newcommand\mpssq{\meter\per\second\squared}
\renewcommand\exp[1]{\ensuremath{\mathrm{e}^{#1}}}
\newcommand\SOHO{SOHO\xspace}
\newcommand\SDO{SDO\xspace}
\newcommand\STEREO{STEREO\xspace}
\newcommand\STEREOA{STEREO~A\xspace}
\newcommand\STEREOB{STEREO~B\xspace}
\newcommand\Nabla{\vec{\nabla}} % differential element
\newcommand{\rainbow}{\textit{rain bow}\xspace}
\newcommand{\vproj}{\ensuremath{v_\parallel} }
\newcommand\Halpha{H\ensuremath{\alpha}\xspace}

\newcommand\Aline[1]{\ensuremath{A_\text{\textit{#1}}(s/L_\text{\textit{#1}})}}

\RequirePackage[normalem]{ulem}
\definecolor{PineGreen}{HTML}{008B72}
\definecolor{Maroon}{HTML}{AF3235}

\DTMnewdatestyle{AAPymd}{}
\DTMnewdatestyle{AAPmd}{}
\DTMnewdatestyle{AAPym}{}
\DTMnewtimestyle{hmsUT}{}
\DTMnewtimestyle{hmUT}{}
\DTMsetdatestyle{AAPymd}
\DTMsettimestyle{hmUT}

\definecolor{coloric}{HTML}{DDAA33}
\definecolor{colorcc}{HTML}{BB5566}
\definecolor{colorpr}{HTML}{004488}
\definecolor{colorfa}{HTML}{AAAAAA}
\definecolor{colorno}{HTML}{FFFFFF}
\newcommand{\bsym}{$\blacksquare$}
\newcommand{\bsymwh}{$\square$}
\newcommand{\bic}{{\color{coloric}\bsym}}
\newcommand{\bcc}{{\color{colorcc}\bsym}}
\newcommand{\bpr}{{\color{colorpr}\bsym}}
\newcommand{\bfa}{{\color{colorfa}\bsym}}
\newcommand{\bno}{{\bsymwh}}
\newcommand{\bwh}{{\phantom{\bsym}}}

\definecolor{todocolor}{HTML}{8B6800}

\begin{document}

\title{
The role of asymmetries in coronal rain formation during thermal non-equilibrium cycles
}

\titlerunning{Asymmetries and coronal rain in TNE cycles}
\authorrunning{G. Pelouze et al.}
\date{}

% authors {{{

\newcommand{\orcid}[1]{}
\author{%
{Gabriel Pelouze}\inst{\ref{aff:CmPA},\ref{aff:IAS}}\orcid{0000-0002-0397-2214}
\and {Frédéric Auchère}\inst{\ref{aff:IAS}} \orcid{0000-0003-0972-7022}
\and {Karine Bocchialini}\inst{\ref{aff:IAS}} \orcid{0000-0001-9426-8558}
\and {Clara Froment}\inst{\ref{aff:LPC2E}} \orcid{0000-0001-5315-2890}
\and {Zoran Mikić}\inst{\ref{aff:PredSci}} \orcid{0000-0002-3164-930X}
\and {Elie Soubrié}\inst{\ref{aff:IAS},\ref{aff:IACCC}} \orcid{0000-0001-9295-1863}
\and {Alfred Voyeux}\inst{\ref{aff:IAS}}
}

\institute{%
\label{aff:CmPA}{Centre for mathematical Plasma Astrophysics, Department of Mathematics, KU Leuven, Celestijnenlaan 200B bus 2400, 3001 Leuven, Belgium.}
\\ \email{gabriel.pelouze@kuleuven.be}
\and \label{aff:IAS}{Institut d'Astrophysique Spatiale, CNRS, Univ. Paris-Sud, Université Paris-Saclay, Bât. 121, 91405 Orsay cedex, France}
\and \label{aff:LPC2E}{LPC2E, CNRS/University of Orl\'eans/CNES, 3A avenue de la Recherche Scientifique, Orl\'eans, France}
\and \label{aff:PredSci}{Predictive Science, Inc., San Diego, CA 92121, USA}
\and \label{aff:IACCC}{Institute of Applied Computing \& Community Code, Universitat de les Illes Balears, 07122 Palma de Mallorca, Spain}
}

% }}}

\abstract%
% (context)
{
Thermal non-equilibrium (TNE) produces several observables that can be used to constrain the spatial and temporal distribution of solar coronal heating.
Its manifestations include prominence formation, coronal rain, and long-period intensity pulsations in coronal loops.
The recent observation of abundant periodic coronal rain associated with intensity pulsations by Auchère et al. allows to unify these two phenomena as the result of TNE condensation and evaporation cycles.
On the other hand, many intensity pulsation events observed by Froment et al. show little to no coronal rain formation. 
}
% aims
{
Our goal is to understand why some TNE cycles produce such abundant coronal rain, while others produce little to no rain.
}
% methods
{
We reconstruct the geometry of the event reported by Auchère et al., using images from STEREO/SECCHI/EUVI and magnetograms from SDO/HMI.
We then perform 1D hydrodynamic simulations of this event, for different heating parameters and variations of the loop geometry (9000 simulations in total).
We compare the resulting behaviour to simulations of TNE cycles by Froment et al. that do not produce coronal rain.
}
% results
{
Our simulations show that both prominences and TNE cycles (with and without coronal rain) can form within the same magnetic structure.
We show that the formation of coronal rain during TNE cycles depends on the asymmetry of the loop and of the heating.
Asymmetric loops are overall less likely to produce coronal rain, regardless of the heating.
In symmetric loops, coronal rain forms when the heating is also symmetric.
In asymmetric loops, rain forms only when the heating compensates the asymmetry.
}
% (conclusions)
{}

\keywords{Sun: atmosphere -- Sun: corona -- Sun: oscillations -- Sun: UV radiation}

\renewcommand{\figureautorefname}{Fig.}
\renewcommand{\sectionautorefname}{Sect.}
\renewcommand{\subsectionautorefname}{Sect.}
\renewcommand{\equationautorefname}{Eq.}

\maketitle

\section{Introduction}
\label{sec:introduction}

Understanding the heating of coronal loops has been a long-standing problem in solar physics.
Numerous physical mechanisms have been proposed \citep[see][and references therein]{CranmerWinebarger2019}.
The main challenge is to identify which ones are predominant in the solar atmosphere.
This requires strong observational diagnostics to discriminate between candidate mechanisms.
To that end, it is important to constrain the spatial and temporal properties of the heating.

Thermal non-equilibrium (TNE) provides several valuable observables to constrain coronal heating.
In the solar corona, TNE occurs in coronal loops which are subject to a highly stratified heating localised close to the surface \citep{AntiochosKlimchuk1991, AntiochosEtAl1999, AntiochosEtAl2000, KarpenEtAl2001, KlimchukLuna2019, Klimchuk2019, Antolin2020}.
In this configuration, the heating may not balance the radiative losses at the apex of the loop.
Such loops thus have no thermal equilibrium state.
As a result of the footpoint heating, chromospheric plasma evaporates, and upflows develop in both legs of the loop.
Plasma accumulates at the loop apex, leading to an increase of the density.
The radiative losses increase quadratically with the density, which at some point, since the heat input is constant, results in an inevitable cooling.
As the temperature drops, the radiative losses further increase \citep{Pottasch1965, McWhirterEtAl1975}, and the plasma quickly starts to condense close to the apex of the loop.
Depending on the geometry of the magnetic field, the condensation may accumulate at coronal heights and form a prominence \citep[\eg][]{MokEtAl1990, AntiochosKlimchuk1991, AntiochosEtAl1999, AntiochosEtAl2000, KarpenEtAl2001, KarpenEtAl2005, XiaEtAl2011, XiaEtAl2014, XiaKeppens2016}.
Alternatively, the condensation may be evacuated from the loop, and fall towards one of its footpoints.
In this latter case, the process starts over, resulting in condensation and evaporation cycles, also referred to as TNE cycles \citep{KuinMartens1982, MartensKuin1983, CraigSchulkes1985}.
The characteristics of TNE cycles (\eg its period, or the minimum and maximum temperature) depend on the spatial and temporal properties of the heating, as well as the geometry of the loop.
These cycles thus constitute a valuable tool to constrain the heating of coronal loops.

TNE cycles result in long-period (\num{2} to \SI{16}{\hour}) pulsations of the extreme-ultraviolet (EUV) emission of coronal loops.
Such pulsations were first detected by \citet{AuchereEtAl2014}
in images from the \textit{Extreme-ultraviolet Imaging Telescope} \citep[EIT;][]{DelaboudiniereEtAl1995} onboard the \textit{Solar and Heliospheric Observatory} \citep[\SOHO;][]{DomingoEtAl1995},
and by \citet{FromentEtAl2015}
in images from the \textit{Atmospheric Imaging Assembly} \citep[AIA;][]{LemenEtAl2012, BoernerEtAl2012} onboard the \textit{Solar Dynamics Observatory} \citep[\SDO;][]{PesnellEtAl2012}.
These long-period intensity pulsations have been found to be very common in coronal loops \citep{AuchereEtAl2014, Froment2016}.
This provides a strong indication that a subset of coronal loops are subject to a stratified heating, localised close to their footpoints.

Another key consequence of TNE is the formation of coronal rain.
It occurs when plasma at coronal heights condenses to chromospheric temperatures before being evacuated \citep{Schrijver2001, MullerEtAl2003, MullerEtAl2004, MullerEtAl2005, DeGroofEtAl2005, AntolinEtAl2010}.
This condensed plasma falls along the coronal loop, forming blob-like structures that are observed in chromospheric and transition-region lines \citep{Kawaguchi1970, Leroy1972, Foukal1978, Schrijver2001, OSheaEtAl2007, DeGroofEtAl2004, DeGroofEtAl2005, AntolinEtAl2010, AntolinEtAl2012, AntolinRouppevanderVoort2012, VashalomidzeEtAl2015}.
Coronal rain can also occur in post-flare loops as a result of the intense transient heating from the flare \citep{ScullionEtAl2016}.
While the formation of periodic coronal rain during TNE cycles had long been suspected \citep[\eg][]{AntolinEtAl2010}, it proved challenging to observe.

The first case of periodic coronal rain associated with EUV intensity pulsations was reported by \citet{AuchereEtAl2018}.
This event, nicknamed the “\rainbow”, was observed in long trans-equatorial loops observed off-limb by \SDO/AIA.
The authors report the detection of EUV intensity pulsations from plasma at coronal temperatures, accompanied by periodic coronal rain observed in the \SI{304}{\angstrom} channel.
The observation in the \SI{304}{\angstrom} channel indicates that the condensations cool down to at least \SI{90000}{K} during the TNE cycle.
The rain is also seen in images of the \Halpha coronograph of the Pic du Midi observatory \citep{KoechlinEtAl2019}, bringing the temperature range all the way down to chromospheric (this was not published by \citealp{AuchereEtAl2018}).
Finally, at the end of the observation sequence, the coronal rain remains longer at the apex of the loop before falling, looking like the onset formation of a prominence.
This event confirms that coronal rain and EUV intensity pulsations are indeed two aspects of TNE cycles.
It also hints to the fact that prominences could result from extremely similar heating or magnetic field configurations.
Further observations of coronal rain associated with EUV intensity pulsations were performed by \citet{FromentEtAl2020}.

On the other hand, TNE cycles can occur with little to no formation of coronal rain.
This happens when the condensing plasma is reheated before it has had time to cool down to chromospheric temperatures.
Entirely ruling out the presence of coronal rain in observations of TNE cycles is difficult because the finer rain drops may not be resolved by the available instruments.
However, \citet{FromentEtAl2015} performed a detailed analysis of the thermal evolution of three EUV intensity pulsation events, and concluded that some of the observed cycles were unlikely to produce coronal rain.
This was later confirmed by the analysis of off-disk observation of these events by the \textit{Extreme Ultraviolet Imager} \citep[EUVI;][]{WuelserEtAl2004}
from the \textit{Sun Earth Connection Coronal and Heliospheric Investigation} instrument suite \citep[SECCHI;][]{HowardEtAl2008}
onboard the \textit{Solar Terrestrial Relations Observatory} \citep[\STEREO;][]{DriesmanEtAl2008, KaiserEtAl2008}.
Finally, flows develop during all TNE cycles, even if the plasma remains at coronal temperature throughout the cycle and no condensation forms \citep{MikicEtAl2013}.
The observation of such flows of plasma at coronal temperature was reported by \citet{PelouzeEtAl2020flows}.

The different consequences of TNE in coronal loops have also been investigated in a number of simulation studies.
The first simulation studies of TNE focused on the evolution of plasma inside a fixed coronal loop, as a response to an imposed heating.
A common approach is to reduce a 3D magnetohydrodynamic (MHD) problem to a 1D hydrodynamic problem, where the plasma evolves along a fixed magnetic field line.
Such simulations produced both TNE cycles with formation of coronal rain \citep{KuinMartens1982, MartensKuin1983, KarpenEtAl2001, KarpenEtAl2005, MullerEtAl2003, MullerEtAl2004, MullerEtAl2005, AntolinEtAl2010, XiaEtAl2011}, and prominences \citep{AntiochosKlimchuk1991, AntiochosEtAl1999, AntiochosEtAl2000, KarpenEtAl2001, KarpenEtAl2005, KarpenEtAl2006, KarpenAntiochos2008, XiaEtAl2011}.
However, simulations by \citet{MikicEtAl2013} showed that the characteristics of TNE cycles depend more strongly on the geometry of the loop than assumed in previous works.
In particular, they demonstrated that TNE cycles without formation of coronal rain (incomplete condensations) could develop in loops with a non-uniform cross-sectional area.
\citet{FromentEtAl2017} later performed simulations that accurately reproduced the TNE cycles with little coronal rain observed by \citet{FromentEtAl2015}.
\citet{FromentEtAl2018} then investigated how the heating and the geometry of the loop interact to produce TNE cycles.
To that end, they simulated the response of three loops of different geometries to a range of heating functions.
They concluded that TNE cycles only occur for a fraction of the explored heating parameters, and require a specific match between the loop geometry and the heating function.

However, it is still unclear why some TNE cycles produce little to no coronal rain \citep[\eg][]{FromentEtAl2015}, while some produce abundant coronal rain \citep[\eg the \rainbow,][]{AuchereEtAl2018}.
Furthermore, the \rainbow forms very abundant rain showers compared to the other periodic coronal rain event \citep{FromentEtAl2020}, and the observations even suggests the onset formation of a prominence.
\citet{FromentEtAl2018} have shown that cycles with and without coronal rain can occur within the same loop, given different heating functions.
Our goal is to better understand which geometry and heating parameters lead to the formation of coronal rain during TNE cycles, and to the extreme behaviour of the \rainbow.
To that end, we perform numerical simulations of the \rainbow event \citep{AuchereEtAl2018}.
We derive eight variants of the geometry, and study how the plasma evolves in each loop as a response to a large number of heating functions.
We then compare the results to the parametric study realised by \citet{FromentEtAl2018} for different loop geometries, in particular that of a loop experiencing TNE cycles with little coronal rain.
This means comparing an event with abundant periodic coronal rain forming in trans-equatorial loops (the \rainbow), to TNE cycles with little to no rain occurring in long loops located at the edge of an active region \citep[][loop~B]{FromentEtAl2018}.

We first determine the geometry of the \rainbow, which is required to perform the simulations (\autoref{sec:rainbow_geometry}).
We then present the numerical setup (\autoref{subsec:sim_numerical_setup}), and the exploration of the geometries and heating parameter space (\autoref{subsec:sim_param_space_scan}).
This parametric study shows that all evolution scenarios of TNE (pulsation at coronal temperatures, periodic coronal rain, and prominence formation) can occur in the \rainbow loop, depending on the heating function (\autoref{subsec:sim_results}).
Additionally, we estimate the spatial distribution of the heating in the observed loops (\autoref{subsec:sim_rainbow_heating_and_expansion}).
We then compare our simulations to the parametric study performed by \citet{FromentEtAl2018} for different loop geometries, in order to understand how the asymmetry of the loop and the heating result in the formation of coronal rain during TNE cycles (\autoref{sec:role_of_asymmetries}).
Finally, we summarise our conclusions in \autoref{sec:summary}.

\section{Geometry of the “\rainbow”}
\label{sec:rainbow_geometry}

\begin{figure*}
\includegraphics[width=\textwidth]{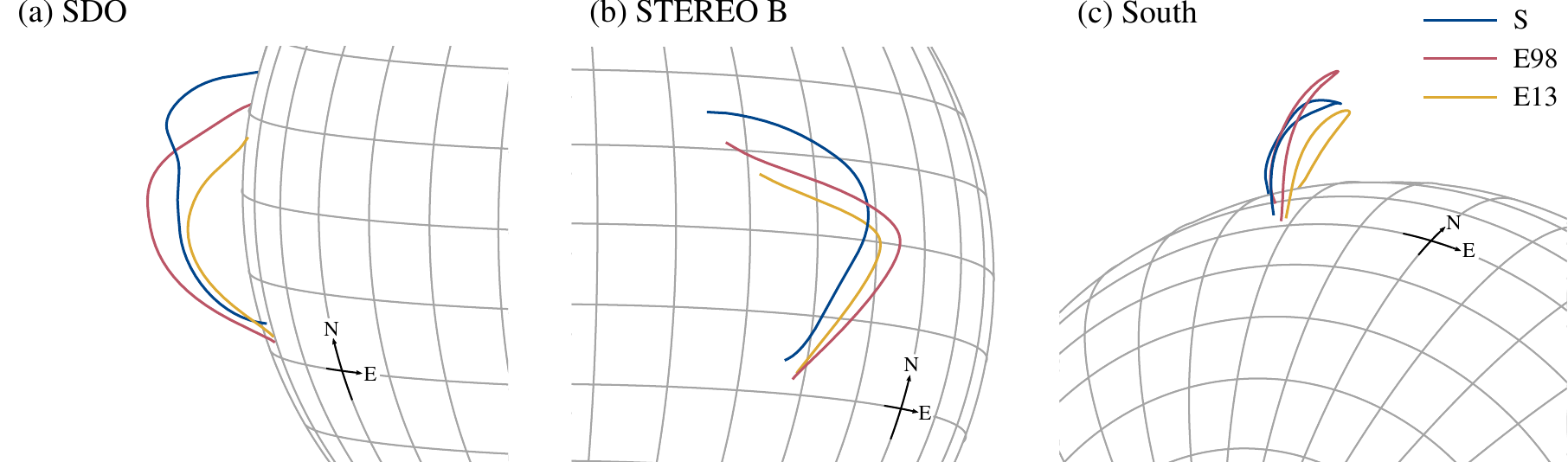}
\caption{
Geometries of the “\rainbow” from the stereoscopic reconstruction (\textit{S}) and from the magnetic field extrapolation (\textit{E13} and \textit{E98}), viewed from the position of \SDO~(a), \STEREOB~(b), and near the solar South pole~(c).
The heliographic grid is spaced by \SI{10}{\degree}.
The cross indicating the North and East directions on the solar sphere is located at the latitude of \SI{-20}{\degree} and the Carrington longitude of \SI{220}{\degree}.
}
\label{fig:rainbow_3d}
\end{figure*}

\begin{figure*}
\includegraphics[width=\textwidth]{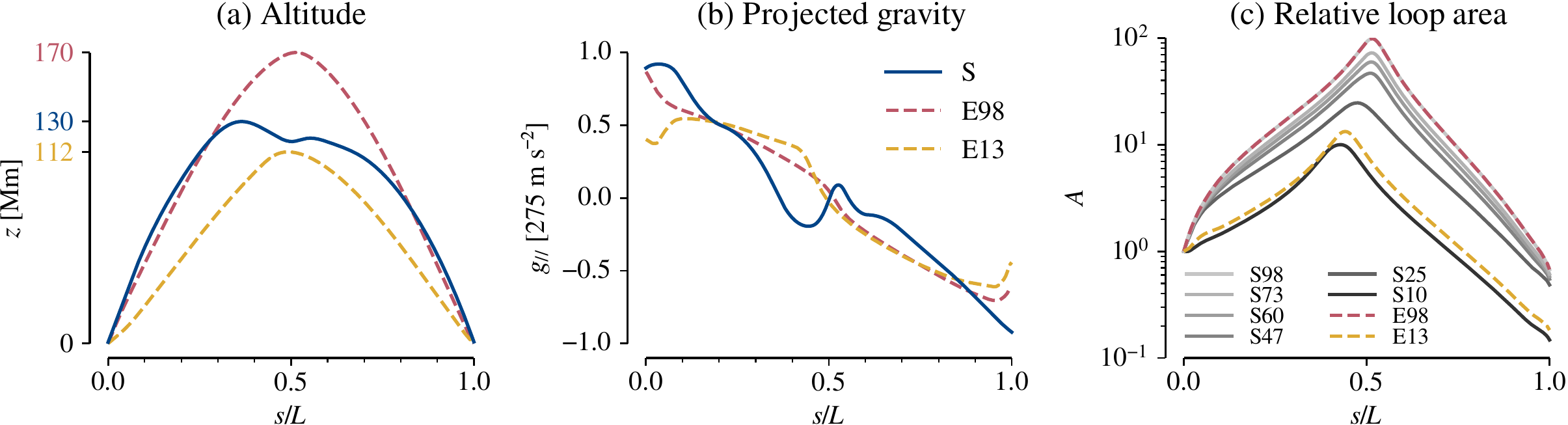}
\caption{
Altitude~(a), projected gravity~(b), and relative loop cross-sectional area~(c)
as a function of the position along the loop $s$ divided by the loop length $L$,
for the geometries shown in \autoref{fig:rainbow_3d}, and the six loop area profiles generated for the $S$ geometry.
The maximum values of the altitude and relative loop area are marked on the corresponding y-axes.
The position $s = 0$ corresponds to the North footpoint of each loop.
}
\label{fig:rainbow_2d}
\end{figure*}

To perform numerical simulations of the \rainbow event \citep{AuchereEtAl2018}, it is necessary to know the 3D geometry of the loop.
For loops observed on-disk, the geometry can be reconstructed using a magnetic field extrapolation \citep[\eg][]{MikicEtAl2013, FromentEtAl2017}.
However, this is not possible for loops observed off-limb such as the \rainbow, as there exist no cotemporal magnetogram.
The closest magnetograms suitable for extrapolation are necessarily out of date, having been taken when the loop was observed on-disk from Earth.
In the case of the \rainbow which is observed at the East limb, that is either at the previous solar rotation (about 20 days before), or once the structure reached the disk centre (about six days later).

However, the \rainbow was also observed on-disk by SECCHI/EUVI onboard \STEREOB.
We combine these observations with images from \SDO/AIA to get a stereoscopic view of the loop and reconstruct its geometry (\autoref{subsec:geometry_stereo} and \autoref{appendix:geometry_stereo}).

Then, we use synoptic magnetograms from the \textit{Helioseismic and Magnetic Imager} \citep[HMI;][]{ScherrerEtAl2012} onboard \SDO to get an estimation of the magnetic field expansion in the loop (\autoref{subsec:geometry_extrapolation} and \autoref{appendix:geometry_extrapolation}).

\subsection{Stereoscopic reconstruction}
\label{subsec:geometry_stereo}

We use stereoscopic observations of \SDO/AIA (near Earth), and \STEREOB/SECCHI/EUVI (separation of \ang{115;;} with Earth) to reconstruct the tridimensional geometry of the \rainbow loop.
The reconstruction is summarised in \autoref{appendix:geometry_stereo}.
The resulting 3D loop shape is shown in \autoref{fig:rainbow_3d} (line \textit{S}).
It has a length of \SI{635.36}{Mm}, and is oriented in the North-South direction.
While it appears to be in a plane when seen from \SDO (\autoref{fig:rainbow_3d}~a), the loop is clearly non-planar and leans towards the East of the solar sphere.
As a result, it is much longer than it appears to be in AIA images (\citealp{AuchereEtAl2018} report a length of \SI{438}{Mm} in the plane of the sky).
\autoref{fig:rainbow_2d}~(a) and (b) show the altitude $z$ and the projected gravity $g_\parallel$ as a function of the position along the loop $s$ divided by the loop length $L$ ($s = 0$ corresponds to the North footpoint of the loop).
These plots reveal the presence of a dip near the apex, which is an unusual feature for a coronal loop.
The dip has a depth of \SI{2}{Mm}, and a width of \SI{53}{Mm}.
The loop legs are otherwise rather symmetric, both having comparable altitude profiles.
In addition, both footpoints are almost vertical, with a projected gravity equal to the solar surface gravity. 

The stereoscopic reconstruction has a few limitations.
The first one is the that loop footpoints are not well constrained.
The second limitation is that the geometry was only reconstructed at a given time, and therefore does not reflect the possible time evolution of the magnetic field.
Finally, it only provides us with the shape of the loop, but not its cross-sectional area.

\subsection{Magnetic field extrapolation}
\label{subsec:geometry_extrapolation}

The hydrodynamic simulations also need as an input the relative cross-sectional loop area.
The relative loop area $A$ is inversely proportional to the magnetic field strength $B$: $A(s) = B(0) / B(s)$.
We normalise the cross-sectional area to the area of the North footpoint (\ie $A(0) = 1$).
To estimate the loop expansion of the \rainbow, we perform a potential magnetic field extrapolation from a low-resolution synoptic line-of-sight magnetogram constructed using \SDO/HMI data.
The construction of the synoptic magnetogram and the magnetic extrapolation are detailed in \autoref{appendix:geometry_extrapolation}.

We select two magnetic field lines extrapolated from this magnetogram.
The shape of these lines is shown in \autoref{fig:rainbow_3d} (lines \textit{E13} and \textit{E98}).
The altitude, projected gravity, and relative loop area are shown in \autoref{fig:rainbow_2d}.
Both loops are non-planar and inclined towards the East (albeit less than the stereoscopically-reconstructed shape \textit{S}), and have no apex dip.
Line \textit{E98} has a length of \SI{628.86}{Mm}, and is overall similar to the stereoscopic shape \textit{S} with no apex dip.
The dip present in shape \textit{S} could result from the deformation of the magnetic field by the abundant coronal rain, because the stereoscopic reconstruction was performed late in the observing sequence, after a lot of rain had already formed. 
Such deformation would not be visible in the result of the potential extrapolation.
Line \textit{E13} is shorter, with a length of \SI{465.54}{Mm}.

As discussed in the appendix, the potential extrapolation yields a very coarse estimation of the magnetic field.
To mitigate this, we build six cross-sectional area profiles from four extrapolated magnetic field lines, and combine them with the shape from the stereoscopic reconstruction \textit{S}.
Thus, we can later verify whether the simulation results strongly depend on the loop area expansion, and assess the importance of the uncertainties associated with the extrapolation.
We label the resulting geometries with their maximum relative area expansion: \textit{S10}, \textit{S25}, \textit{S47}, \textit{S60}, \textit{S73}, and \textit{S98}.
The cross-sectional area profiles of these geometries are shown in \autoref{fig:rainbow_2d}~(c).

We perform numerical simulations for the six geometries using the stereoscopic loop shape, as well as for the two extrapolated geometries.

\section{Numerical simulations of TNE cycles}
\label{sec:sim_rainbow}

\subsection{Numerical setup}
\label{subsec:sim_numerical_setup}

We use the 1D hydrodynamic setup described by \citet{MikicEtAl2013}.
This setup allows us to compute the evolution of the plasma inside a fixed magnetic field tube of non-uniform cross-sectional area.
To that end, the equations for the conservation of mass, momentum, and energy are solved.
The energy equation includes radiative losses, thermal conduction, and a constant, non-uniform imposed heating term $H(s)$.
The radiative losses are computed using the Chianti atomic database \citep[version 7.1;][]{DereEtAl1997-chianti1, LandiEtAl2013-chianti13v7-1}, coronal abundances \citep{FeldmanEtAl1992, Feldman1992, GrevesseSauval1998, LandiEtAl2002-chianti5}, and the Chianti collisional ionization equilibrium model \citep{DereEtAl2009-chianti9}.
The thermal conduction is computed using the Spitzer conductivity $\kappa_\parallel(T)$ \citep{Spitzer1962} above the cut-off temperature $T_c = \SI{250000}{K}$, and a constant conductivity $\kappa_\parallel(T_c)$ at lower temperatures \citep{LionelloEtAl2009, MikicEtAl2013}.
This artificially broadens the transition region, allowing for larger numerical grid cells and thus a significant reduction of the computation time.
\citet{MikicEtAl2013} showed that the use of the modified conductivity, as well as the choice of the radiative loss function and of the abundances, do not significantly change the simulation results.
In this setup, it is assumed that the plasma is composed of fully ionised hydrogen (thus the electron and ion number densities are equal), and that the ion and electron temperatures are equal.

The loop is initialised in hydrostatic equilibrium, with a parabolic temperature profile reaching \SI{2}{MK} at the loop midpoint, and a \SI{3.5}{Mm}-thick chromosphere at a constant temperature of \SI{20000}{K}.
Throughout the simulation, the constant chromospheric temperature $T_\mathrm{ch}$ and number density $N_\mathrm{ch}$ are imposed at the domain boundaries ($s = 0$ and $s = L$).
We set $T_\mathrm{ch} = \SI{20000}{K}$ and $N_\mathrm{ch} = \SI{3e19}{m^{-3}}$.
This boundary condition allows us to maintain a layer of chromospheric plasma at both footpoints, which serves as a mass reservoir for evaporation, and absorb falling condensations.
Previous studies using this numerical setup imposed $N_\mathrm{ch} = \SI{6e18}{m^{-3}}$ \citep{MikicEtAl2013, DownsEtAl2016, FromentEtAl2017, FromentEtAl2018}.
However, some simulations of the \rainbow produce very large condensations that are not absorbed by the resulting chromospheric layer.
Such condensations cause numerical issues that interrupt the simulation when they reach the boundaries of the computation domain.
To reduce the number of interrupted simulations, we increased the chromospheric density to $\SI{3e19}{m^{-3}}$, thus increasing the thickness of the chromospheric layer.
While this value may be larger than typical chromospheric densities, it does not affect the solution in the coronal part of the loop \citep[see discussion in][]{MokEtAl2005}.
We verified that the behaviour in the coronal part of the simulation did not change when using $N_\mathrm{ch} = \SI{6e18}{m^{-3}}$ and $\SI{3e19}{m^{-3}}$.

The loop is subject to a constant stratified heating, localised near both footpoints.
The heating consists of a uniform background term, and two exponentially decreasing terms, one for each leg of the loop.
It is given by:

\begin{equation}\label{eqn:heating_function}
H(s) = H_0 + H_1 \left( \exp{-g(s)/\lambda_1} + \exp{-g(L-s)/\lambda_2} \right)
\end{equation}

\noindent
where $H_0$ is the constant background volumetric heating rate,
$H_1$ the stratified volumetric heating rate,
and $\lambda_1$ and $\lambda_2$ are the scale heights of the stratified heating at both footpoints.
The term $g(s) = \max(s - \Delta, 0)$ allows for a constant heating in the chromosphere, up to $\Delta = \SI{5}{Mm}$ away from both footpoints.
To obtain a stratified heating, we choose $H_1 \gg H_0$, and $\{\lambda_1, \lambda_2\} < L$.
A greater stratification is obtained with smaller heating scale height values, or higher values of $H_1 / H_0$.
We note that more energy is deposited in the loop as either $\lambda_1$ or $\lambda_2$ increase.
A symmetric heating is obtained when $\lambda_1 = \lambda_2$.

We use \num{10000} grid points along the loop, with variable cell sizes of \SI{4}{km} in the chromosphere, \SI{40}{km} in the transition region, and \SI{400}{km} around the loop apex.
We let the system evolve for \SI{72}{h}.

\subsection{Parameter-space scan: loop geometry and heating}
\label{subsec:sim_param_space_scan}

\begin{table}
\footnotesize
\caption{Explored parameter space:
simulated geometries
with their loop length $L$,
maximum relative area expansion $A_\text{max}$,
and constant background heating $H_0$,
as well as the explored values of the stratified heating intensity $H_1$
and the heating scale heights $\lambda_1$ and $\lambda_2$.
  }
\label{tab:param_space}
\begin{tabular}{lrrrll}
\hline
\hline
  \shortstack{Geometry \\~} &
  \shortstack{$L$ \\ ~[Mm]} &
  \shortstack{$A_\text{max}$ \\~} &
  \shortstack{$H_0$ \\ ~[\si{\nano\watt\per\cubic\meter}]} &
  \shortstack{$H_1 / H_0$ \\~} &
  \shortstack{$\{\lambda_1, \lambda_2\} / L$ \\~} \\
\hline
\textit{S10}
   & \multirow{6}{*}{\makecell{635.36}}
   & \makecell{10.0}
   & \makecell{20.3}
   & \multirow{8}{*}{%
      \begin{minipage}{1cm}
      \raggedright
      Five values each:
      1280, 2560,
      5120, 10240,
      20480.
      \end{minipage}
      }
   & \multirow{8}{*}{%
      \begin{minipage}{1.2cm}
      \raggedright
      \num{15} values each,
      from 1\% to 20\%.
      \end{minipage}
      }
   \\
\textit{S25}  &                   & \makecell{24.6}  & \makecell{19.0}  & & \\
\textit{S47}  &                   & \makecell{46.7}  & \makecell{15.5}  & & \\
\textit{S60}  &                   & \makecell{59.5}  & \makecell{14.5}  & & \\
\textit{S73}  &                   & \makecell{72.5}  & \makecell{14.0}  & & \\
\textit{S98}  &                   & \makecell{98.4}  & \makecell{13.0}  & & \\ \cline{2-2}
\textit{E98}  & \makecell{628.86} & \makecell{98.4}  & \makecell{14.0}  & & \\
\textit{E13}  & \makecell{465.54} & \makecell{13.3}  & \makecell{53.0}  & & \\ \hline
\end{tabular}
\end{table}

We investigate how the evolution of the plasma in the loop depends on the parameters of the heating function (\autoref{eqn:heating_function}), the cross-sectional area of the loop, and the presence of a dip at the apex of the loop.
To that end, we perform simulations for the eight geometries described in \autoref{sec:rainbow_geometry} and shown in Figs.~\ref{fig:rainbow_3d} and~\ref{fig:rainbow_2d}.
Geometries \textit{S10}, \textit{S25}, \textit{S47}, \textit{S60}, \textit{S73}, and \textit{S98} share the stereoscopically reconstructed loop shape, with a length of \SI{635.36}{Mm} and a dip at the apex.
Their maximum area expansions vary from \num{10} to \num{98}.
Geometries \textit{E13} and \textit{E98} are obtained from magnetic extrapolations and do not have apex dips.
Geometry \textit{E98} is the closest to the stereoscopic geometries, with a length of \SI{628.86}{Mm}.
\textit{E13} is shorter, with a length of \SI{465.54}{Mm}.

For each of these geometries, we choose $H_0$ such that the apex temperature is approximately \SI{1}{MK} when the loop is subject to a uniform heating (\ie $H_1 = 0$).
We then choose five values for $H_1$, expressed as multiples of $H_0$: $(1280, 2560, 5120, 10240, 20480) \times H_0$.
We also select 15 values for each heating scale height $\lambda_1$ and $\lambda_2$, ranging from \SI{1}{\percent} to \SI{20}{\percent} of the loop length $L$.
While the most extreme heating functions may no longer correspond to realistic coronal conditions (typically functions with high values of all three parameters), this choice of parameters allows us to not neglect some realistic functions where only one or two parameters reach extreme values.
% I wonder whether this is some kind of scientific FOMO.
In particular, we will show in \autoref{subsec:sim_rainbow_heating_and_expansion} that a high value of $H_1 / H_0$ is required in order to reproduce the observed behaviour of the \rainbow.

We perform simulations for all possible combinations of these heating parameters.
This corresponds to \num{1125} simulations for each loop geometry, for a total of \num{9000} simulations.
The properties of the loop geometries and of the explored heating parameters are summarised in \autoref{tab:param_space}.

\subsection{Results: all TNE consequences in the same simulation set}
\label{subsec:sim_results}

\begin{figure*}
\includegraphics[height=0.90\textheight]{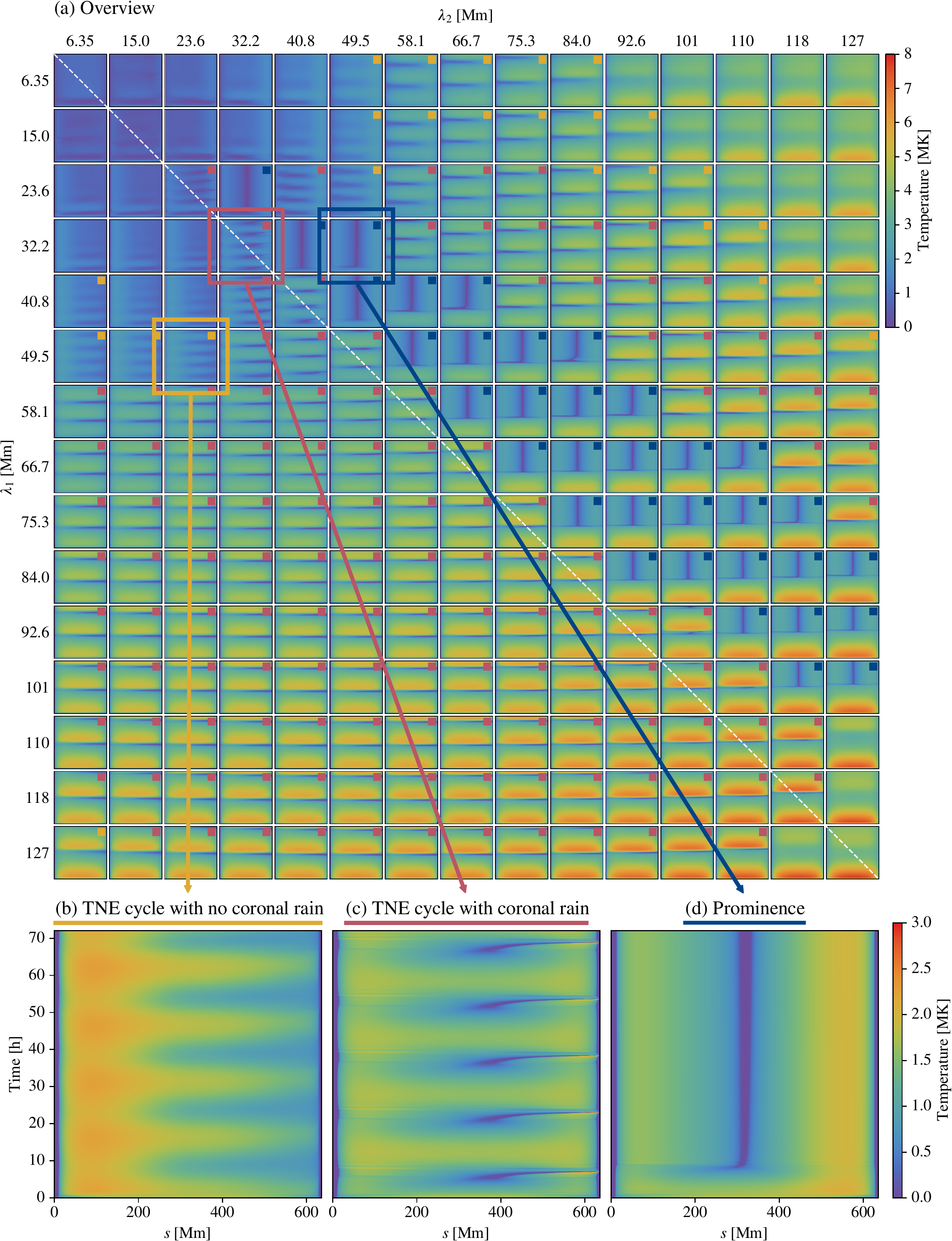}
\caption{
Temperature evolution for the 225 simulations of the \rainbow using the geometry \textit{S47}
(stereoscopically reconstructed shape, with a maximum area expansion $A_\text{max} = 47$),
a constant heating rate $H_0 = \SI{15.5}{\micro\watt\per\cubic\meter}$,
and a stratified heating rate $H_1 = 2560 H_0 = \SI{39.7}{\micro\watt\per\cubic\meter}$.
(a)~Overview of the simulations performed with $\lambda_1$ and $\lambda_2$ each varying from \SI{1}{\percent} to \SI{20}{\percent} of the loop length $L$.
Each sub-plot shows the evolution of temperature as a function of the position along the loop $s$ (x-axis, ranging from \num{0} to $L = \SI{635.36}{Mm}$) and of time (y-axis, ranging from \num{0} to \SI{72}{h}).
Event types are marked by coloured squares: TNE cycles with no coronal rain (\bic), cycles with rain (\bcc), and prominence-like structures (\bpr).
The white dashed line corresponds to $\lambda_1 = \lambda_2$.
(b)~Sample simulation showing TNE cycles without coronal rain.
(c)~Sample simulation showing TNE cycles with coronal rain.
(d)~Sample simulation showing the formation of a prominence-like structure.
}
\label{fig:simu_samples}
\end{figure*}

\begin{figure*}
\includegraphics[height=0.407\textwidth]{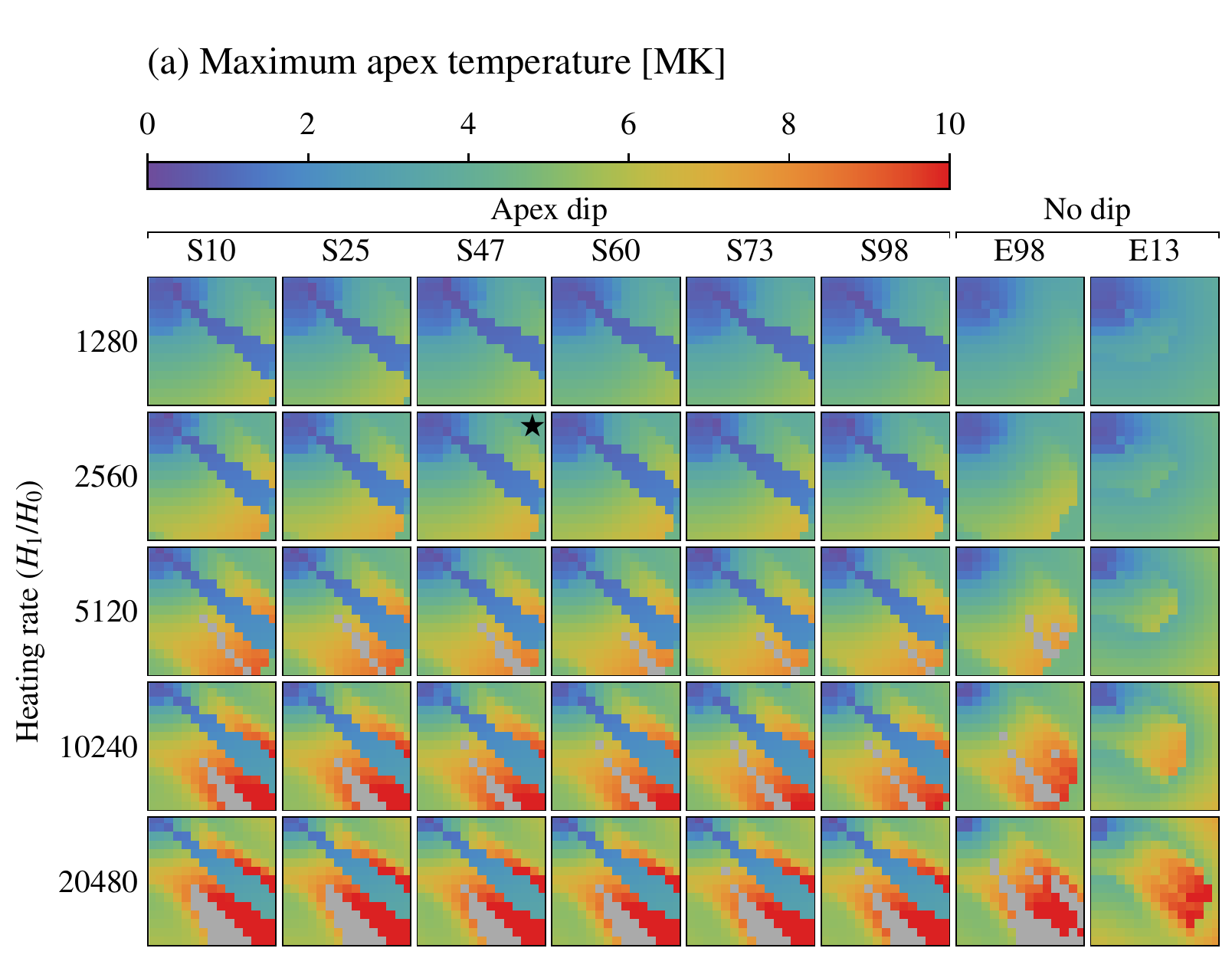}
\includegraphics[height=0.407\textwidth]{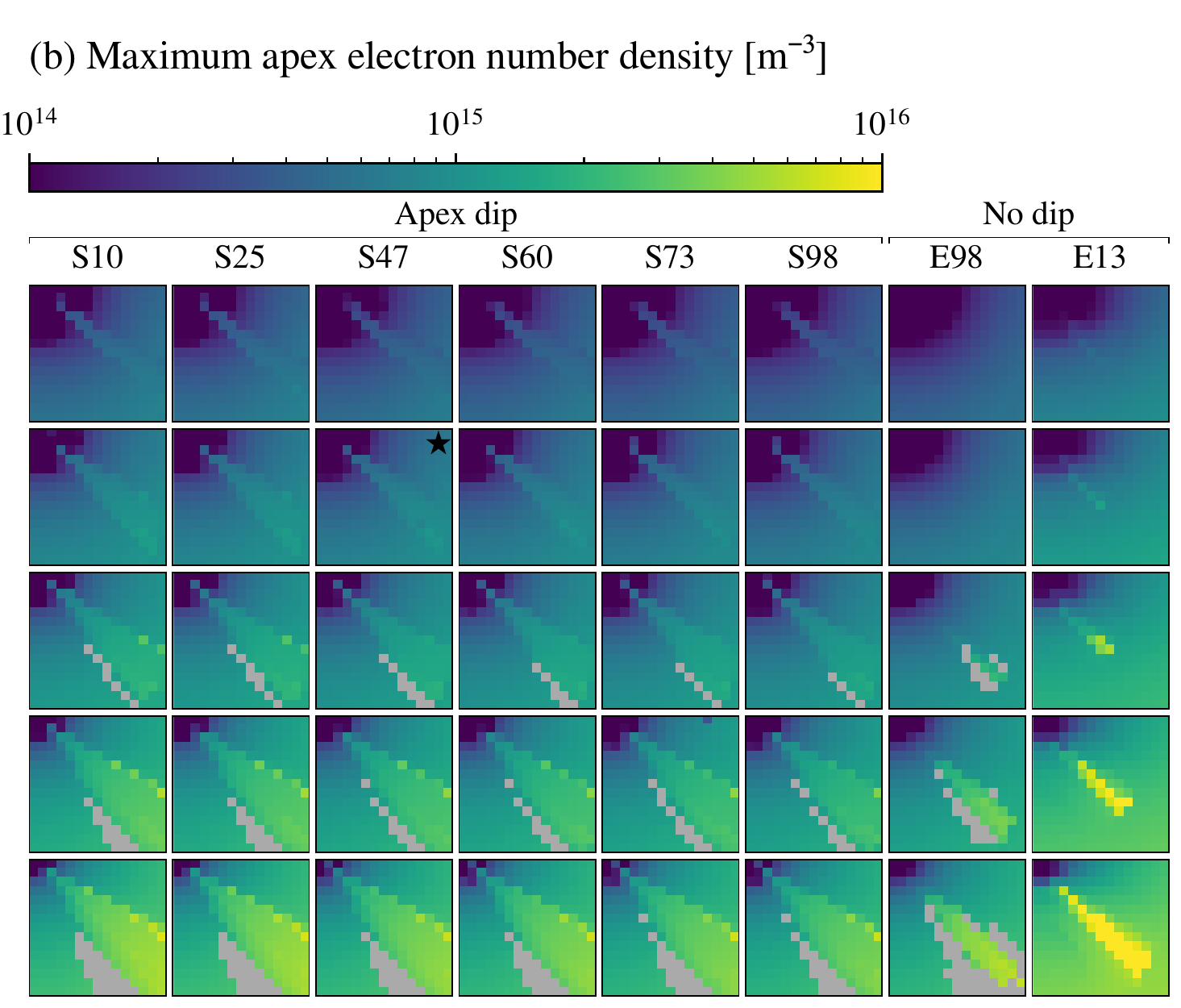}

\includegraphics[height=0.407\textwidth]{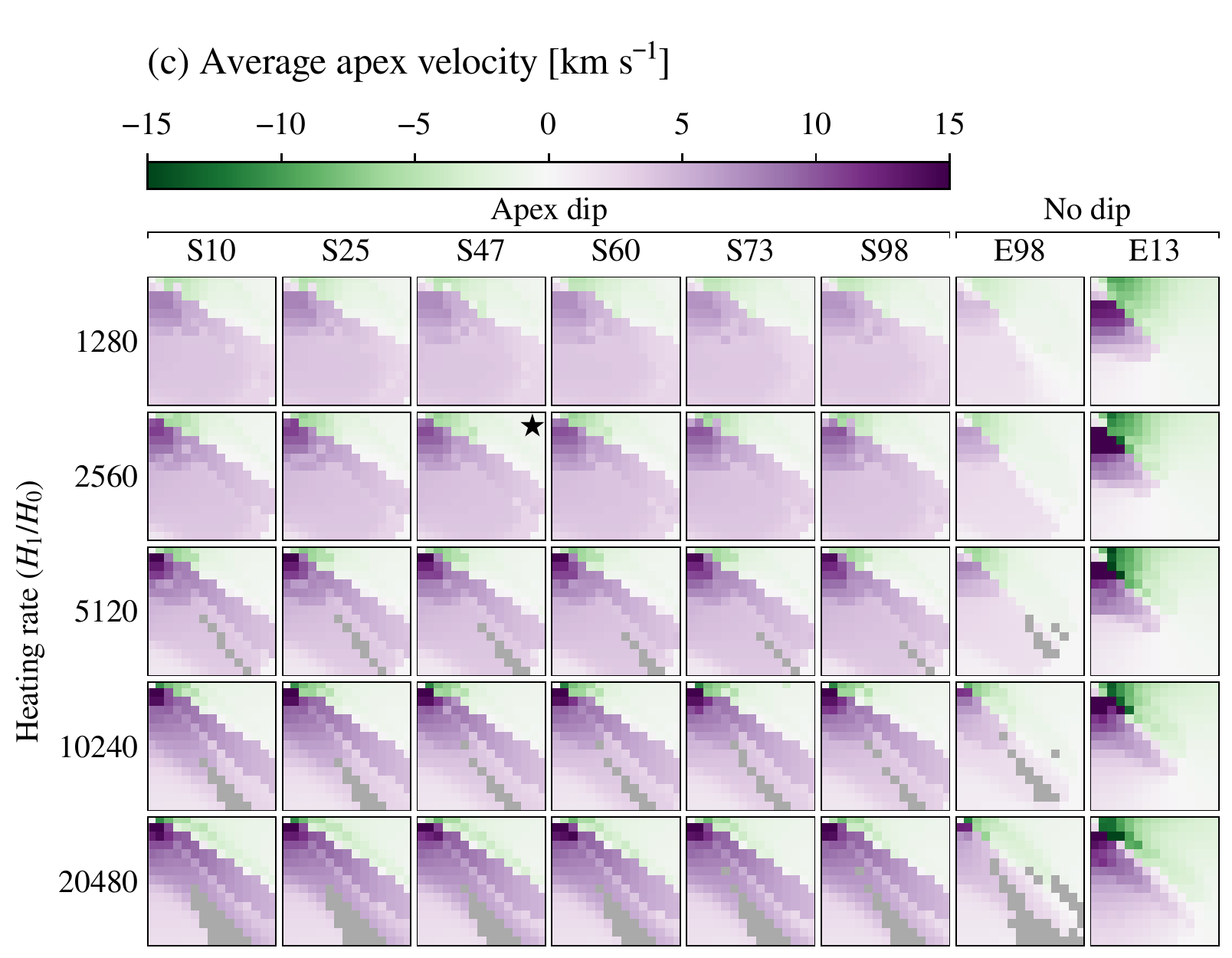}
\includegraphics[height=0.407\textwidth]{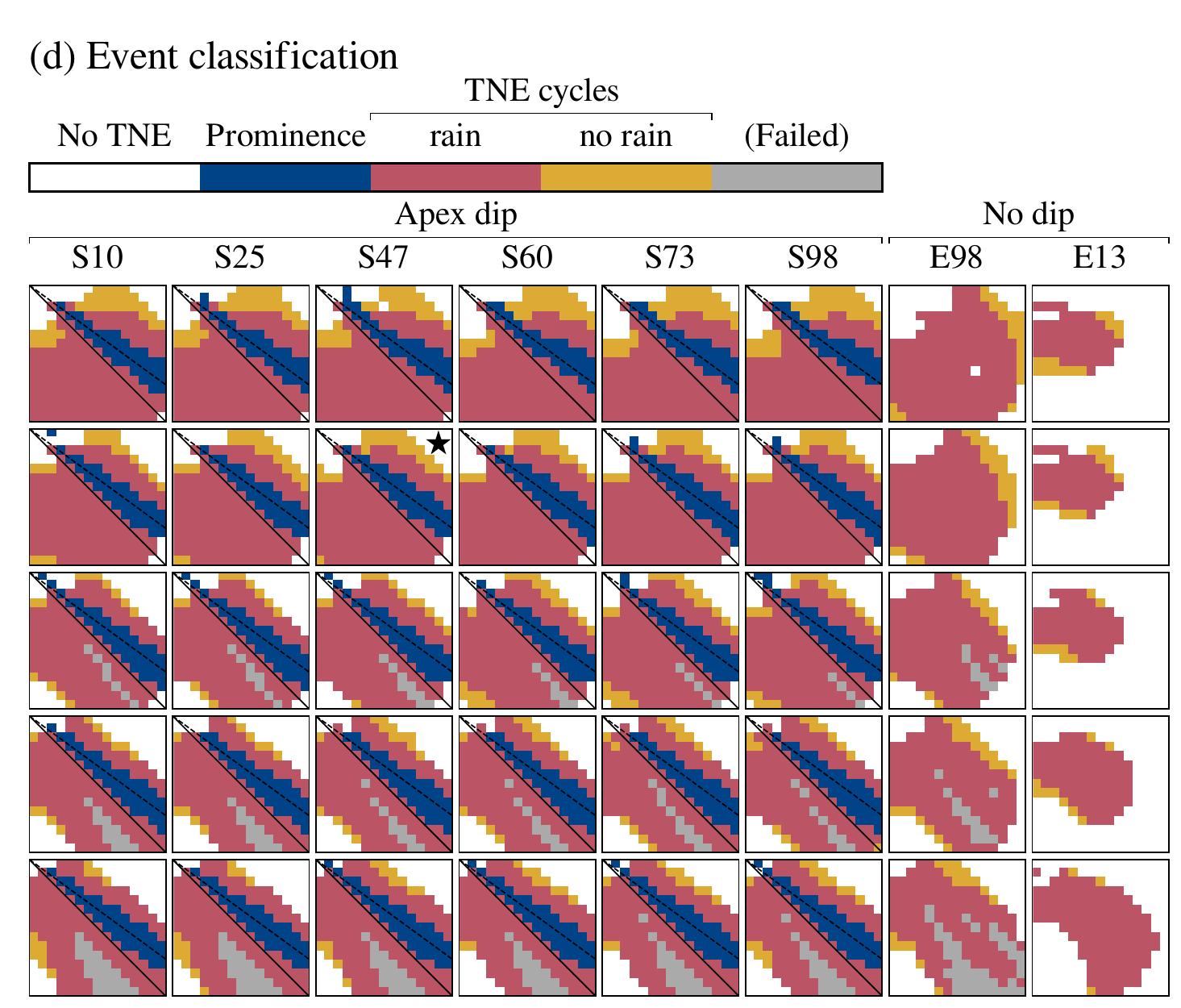}
\caption{
Characteristic quantities for the \num{9000} simulations of the \rainbow.
(a)~Maximum temperature,
(b)~maximum electron number density,
(c)~average velocity,
computed around the loop apex ($z > \num{0.9}z_\mathrm{max}$) during the second half of each simulation ($t > \SI{36}{h}$).
(d)~Event classification: no TNE behaviour (\bno), formation of a prominence-like structure (\bpr), TNE cycles with (\bcc) and without (\bic) coronal rain, and simulations that failed because of numerical issues (\bfa).
Each sub-figure contains 40 squares, corresponding to the eight line geometries (\textit{columns}), and the five values of $H_1 / H_0$ (\textit{rows}).
Each square contains 225 simulations for the different values of $\lambda_1$ (\textit{y-axis}) and $\lambda_2$ (\textit{x-axis}), ranging from \SI{1}{\percent} of the loop length (\textit{top-left}) to \SI{20}{\percent} of the loop length (\textit{bottom-right}).
The \textit{solid lines} on (d) correspond to $\lambda_1 = \lambda_2$.
The \textit{dashed lines} correspond to $\lambda_1 = 0.73 \lambda_2$.
The square marked with a star ($\star$) corresponds to the 225 simulations shown in \autoref{fig:simu_samples}~(a).
}
\label{fig:full_study}
\end{figure*}

For each simulation, we obtain the temperature $T$, electron number density $N_e$, pressure $p$, and velocity along the loop $\vproj$ as a function of time and position along the loop.
In \autoref{fig:simu_samples}~(a), we show the evolution of the simulations performed with the geometry \textit{S47}, a heating rate of $H_1 = 2560 H_0 = \SI{39.7}{\micro\watt\per\cubic\meter}$, and all explored values of the heating scale heights $\lambda_1$ and $\lambda_2$.
This corresponds to 225 of the total 9000 simulations.
Different TNE behaviours are obtained depending on the values of $\lambda_1$ and $\lambda_2$:
\begin{itemize}
\item TNE cycles without formation of coronal rain, where the plasma in the coronal part of the loop remains at coronal temperatures throughout the simulation (\autoref{fig:simu_samples}~b);
\item TNE cycles with formation of coronal rain (\autoref{fig:simu_samples}~c);
\item the formation of a prominence-like structure, where condensed\footnote{The temperature and electron-number density reach chromospheric values, typically $T \sim \SI{0.1}{MK}$ and $N_e \sim \SI{e16}{m^{-3}}$.} plasma remains trapped in the dip at the apex of the loop (\autoref{fig:simu_samples}~d);
\item finally, a few cases with no TNE where the loop reaches a steady state, that is either a static equilibrium or continuous siphon flows (\eg at $\lambda_1 = \SI{6.35}{Mm}, \lambda_2 = \SI{127}{Mm}$).
\end{itemize}

The cycles without coronal rain correspond to the cycles with incomplete condensations (IC) described by \citet{MikicEtAl2013} and \citet{FromentEtAl2018}.
Cycles with coronal rain correspond to cycles with complete condensations (CC).

The beginning of simulations forming a prominence-like structure is very similar to that of simulations showing TNE cycles with coronal rain.
Both begin with an initial relaxation, followed by a plasma condensation.
However, in the case of the prominence-like structures, the condensation remains trapped in the dip at the apex of the loop.
In the case of TNE cycles with coronal rain, the condensation is free to fall along one of the loop legs.

These simulations show the three behaviours that can result from TNE in coronal loops: evaporation and condensation cycles with coronal rain, cycles without coronal rain, and formation of a prominence-like structure.
As the examples in \autoref{fig:simu_samples} show, different events can occur within the same loop geometry, and with similar heating parameters.
Simulations by \citet{MikicEtAl2013}, \citet{MokEtAl2016}, and \citet{FromentEtAl2018} have already shown that cycles with and without coronal rain can occur in the same loop.
However, the current parametric study is, to our best knowledge, the first to show that the three consequences of TNE can occur within the same loop geometry.

\subsubsection{Analysis of simulation results}

To reduce the large amount of simulations, we summarise each simulation to a few characteristic quantities shown in \autoref{fig:full_study}.
The maximum temperature and electron number density reached around the apex of the loop ($z > \num{0.9} z_\mathrm{max}$) during the second half of the simulation ($t > \SI{36}{h}$) are shown in \autoref{fig:full_study}~(a) and (b), respectively.
The velocity averaged at the loop apex during the second half of the simulation is shown in \autoref{fig:full_study}~(c).
Finally, the type of behaviour obtained in each simulation is presented in \autoref{fig:full_study}~(d).

Determining the behaviour in each simulation first requires to identify simulations with TNE cycles, and then those which produce coronal rain.
It also requires to identify the simulations that form prominence-like structures.
\citet{FromentEtAl2018} detected simulations with pulsations in the Fourier space.
However, many of the cycles produced in our simulations have very long periods, ranging from \num{20} to \SI{30}{\hour}.
This gives between \num{2.5} and \num{3.5} periods in a simulation of \SI{72}{\hour}, which is not easily detected in the Fourier space.
We thus identify pulsations manually, by looking at the evolution of the temperature in the loops.
We then apply the same criterion as \citet{FromentEtAl2018} to detect coronal rain:
a simulation is considered to produce coronal rain if the temperature in the coronal part of the loop (defined as $z > \SI{20}{Mm}$) locally drops below \SI{0.5}{MK} after the first \SI{10}{\hour} of the simulation.
Similarly, prominence-like structures are identified manually based on the temperature evolution.

\subsubsection{Stereoscopically reconstructed loop}

We first present the result of the simulations performed with the geometries \textit{S10} to \textit{S98}.
These geometries all use the stereoscopically reconstructed loop shape, combined with different cross-sectional area profiles.

The event classification for all simulations is shown in \autoref{fig:full_study}~(d).
TNE occurs in a large fraction of the explored parameter space, in particular for smaller values of $H_1 / H_0$.
This includes both TNE cycles (with and without coronal rain) and the formation of prominence-like structures.
Prominence-like structures (in blue in \autoref{fig:full_study}~d) form for relatively symmetric heating functions, with scale heights close to the line $\lambda_1 = 0.73 \lambda_2$.
In contrast, TNE cycles with coronal rain (in red) occur for more asymmetric heating functions, and cycles without coronal rain (in yellow) for even more asymmetric functions.
In a small fraction of the parameter space, the loop can reach a steady state, and thus shows no TNE behaviour (in white).
Finally, some simulations fail prematurely because of numerical issues (in grey), and therefore we cannot determine their behaviour.
Theses numerical issues occur when massive condensations fall through the chromospheric layer, and reach a boundary of the computation domain (see \autoref{subsec:sim_numerical_setup}).
Such condensations form for relatively strong and symmetric heating functions.
The strong heating causes important evaporation, while its symmetry allows for the condensations to gain significant mass before falling.

\autoref{fig:full_study}~(a) shows the maximum temperature reached around the loop apex during the second half of each simulation.
Simulations with TNE cycles reach higher temperatures than the neighbouring simulations without cycles, which is consistent with the results of \citet{FromentEtAl2018}.
In addition, loops with a larger area expansion reach slightly lower maximum temperatures (the maximum temperatures reached with the \textit{S98} geometry are approximately \SI{10}{\percent} lower than those reached with the \textit{S10} geometry).
This effect has been explained for static loops by \citet{VeseckyEtAl1979}: a larger area expansion increases the upwards thermal conduction, which is, in turn, compensated by larger radiative losses in the corona.
Because the radiative losses are a decreasing function of the temperature in the corona, this results in lower coronal temperatures.
However, the temperature variation remains small compared to the cross-sectional area variation.
The maximum temperature also increases when the heating rate $H_1 / H_0$ increases.

\autoref{fig:full_study}~(b) shows the maximum electron number density reached around the loop apex during the second half of the simulations.
The maximum electron density is higher in simulations close to the diagonal $\lambda_1 = \lambda_2$.
However, there is no clear-cut relationship between the simulations that produce prominence-like structures, and adjacent simulations that produce TNE cycles with coronal rain.
The maximum electron density increases with the heating rate $H_1 / H_0$ (as expected from loop scaling laws), and slightly decreases with the loop area expansion.
Similarly to the temperature, the density variation is small compared to the cross-sectional area variation (approximately \SI{25}{\percent} difference between \textit{S10} and \textit{S98}).
In some simulations, the plasma $\beta$ (\ie the ratio of gas and magnetic pressures) is greater than one, which breaks the assumption that the magnetic field does not evolve.
This occurs for simulations with large heating rate ($H_1 / H_0$) and scale heights ($\lambda_1$ and $\lambda_2$).
In the most extreme case, the plasma $\beta$ locally reaches \num{500}.
However, such deviations only occur locally, when complete condensations are formed.
As a result, their effect is limited to the local dynamics and morphology of the coronal rain or prominence.
Thus, we do not expect them to have an effect on whether condensations are formed in the first place.
In addition, the actual plasma $\beta$ is likely to be lower than the aforementioned value, because the intensity of the extrapolated magnetic field is probably under-estimated (see discussion in \autoref{appendix:geometry_extrapolation}).
Finally, while beyond the assumptions of the simulation, a high plasma $\beta$ is consistent with the observations.
It would allow the massive condensations to deform the magnetic field and form the observed apex dip.
The gradual formation of the apex dip would also explain why coronal rain is only seen to linger at the end of the sequence observed by \citet{AuchereEtAl2018}.

\autoref{fig:full_study}~(c) shows the average velocity around the loop apex during the second half of the simulations.
For low heating rates $H_1 / H_0$, simulations with prominence-like structures and neighbouring simulations forming TNE cycles display similar apex velocities.
In this case, the average velocities are of the order \SI[retain-explicit-plus]{+5}{\kmps} when $\lambda_1 > 0.6 \lambda_2$ (\ie simulations for which $\lambda_1 > \lambda_2$, plus simulations with prominence-like structures), and of the order of \SI{-5}{\kmps} otherwise.
For higher heating rates, simulations with prominence-like structures have higher average apex velocities than neighbouring simulations with TNE cycles, reaching values of $\sim \SI{8}{\kmps}$.
In simulations that form a prominence-like structure, upflows develop in both legs of the loop and feed material into the prominence, while the velocity in the apex dip is zero.
The North edge of the dip (where velocities are positive) has a higher altitude (see \autoref{fig:rainbow_2d}), and thus contributes more than the South edge to the velocity averages shown in \autoref{fig:full_study}~(c).
This explains why the averaged velocities around the apex are positive in simulations that form prominence-like structures.

\subsubsection{Extrapolated geometries}

We now present the results of the simulations performed with the geometries \textit{E13} and \textit{E98}, obtained through magnetic field extrapolation.
The goal of these simulations is to explore how the presence of the dip at the apex of the stereoscopically-reconstructed plasma affects the evolution of the loops.
Geometry \textit{E98} is similar to geometry \textit{S98}, but with no apex dip:
it has a length of \SI{629}{Mm} (\SI{635}{Mm} for \textit{S98}),
a relative area expansion of \num{98} (identical to that of \textit{S98}),
and a similar altitude profile in the loop legs.
Geometry \textit{E13} has an area expansion of \num{13} (thus comparable to that of \textit{S10}), but a much shorter length (\SI{465.54}{Mm}).

The characteristic values and event classification of the simulations using these geometries are shown in \autoref{fig:full_study}.
Simulations of \textit{E98} mainly differ from those of \textit{S98} in the fact that they do not form prominence-like structures.
This is the direct result of the fact that \textit{E98} has no apex dip.
As a result, all condensations fall along the loop legs, resulting in TNE cycles.
Apart from this, simulations of \textit{E98} and \textit{S98} have overall similar characteristics, including the period of the TNE cycles.
This suggests that the presence of an apex dip leads to the formation of prominence-like structures when the heating is relatively symmetric, but does not otherwise change the characteristics of TNE cycles.
On the other hand, simulations of \textit{E13} have very different characteristics.
This is primarily explained by the fact that geometry \textit{E13} is significantly shorter than the other geometries.

\subsection{Reproducing the observations of the \rainbow}
\label{subsec:sim_rainbow_heating_and_expansion}

\begin{figure}
\includegraphics[width=\columnwidth]{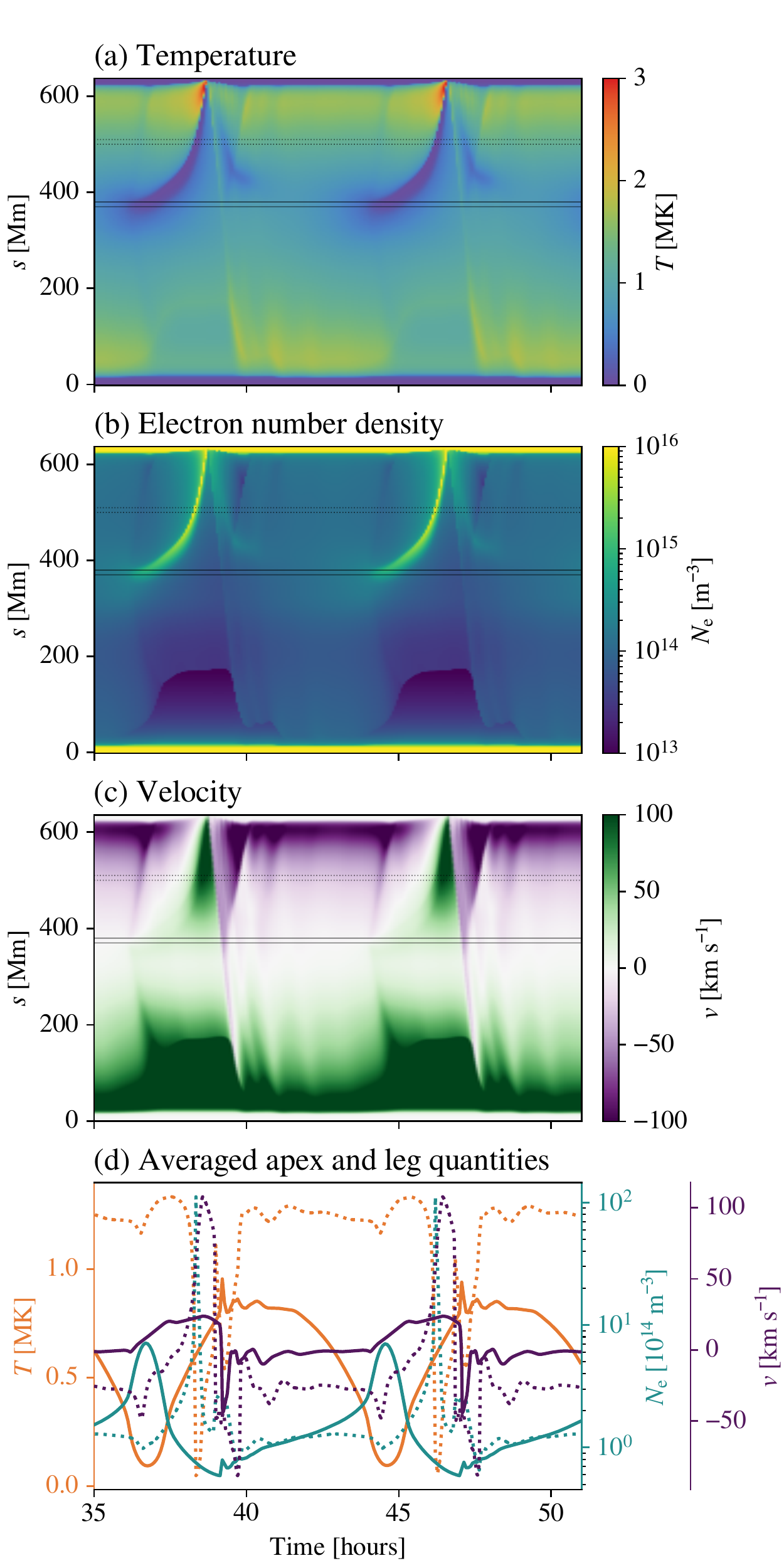}
\caption{
Simulation that best reproduce the observations of the \rainbow,
obtained with geometry \textit{S60}, $H_1 / H_0 = 20480$, and $\lambda_1 = \lambda_2 = \SI{15}{Mm}$.
(a)~Temperature,
(b)~electron number density,
(c)~velocity,
as a function of time and position along the loop.
(d)~Temperature (\textit{orange}), electron number density (\textit{blue}), and velocity (\textit{purple}) averaged
close to the apex ($\SI{370}{Mm} < s < \SI{380}{Mm}$, \textit{solid lines})
and in the leg ($\SI{500}{Mm} < s < \SI{510}{Mm}$, \textit{dashed lines}).
The apex and leg regions are also shown in sub-figures (a), (b), and (c).
The figure shows times from \num{35} to \SI{51}{h} out of the \SI{72}{h} of the simulation.
The temporal evolution of the temperature, electron number density, and velocity along the loop can be seen in the online movie.
}
\label{fig:simu_rainbow_t_ne_v}
\end{figure}

\begin{figure}
\includegraphics[width=\columnwidth]{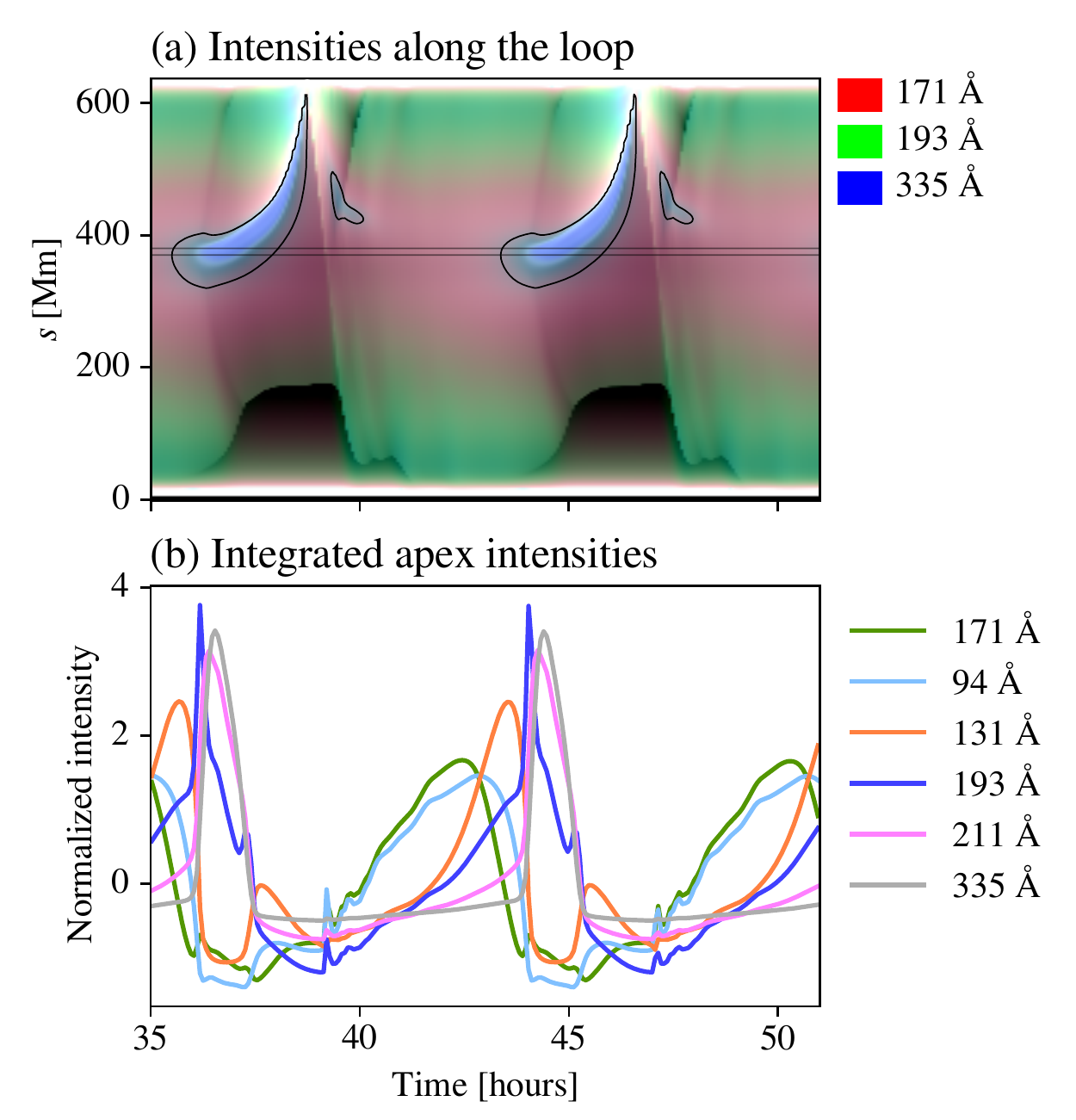}
\includegraphics[width=\columnwidth]{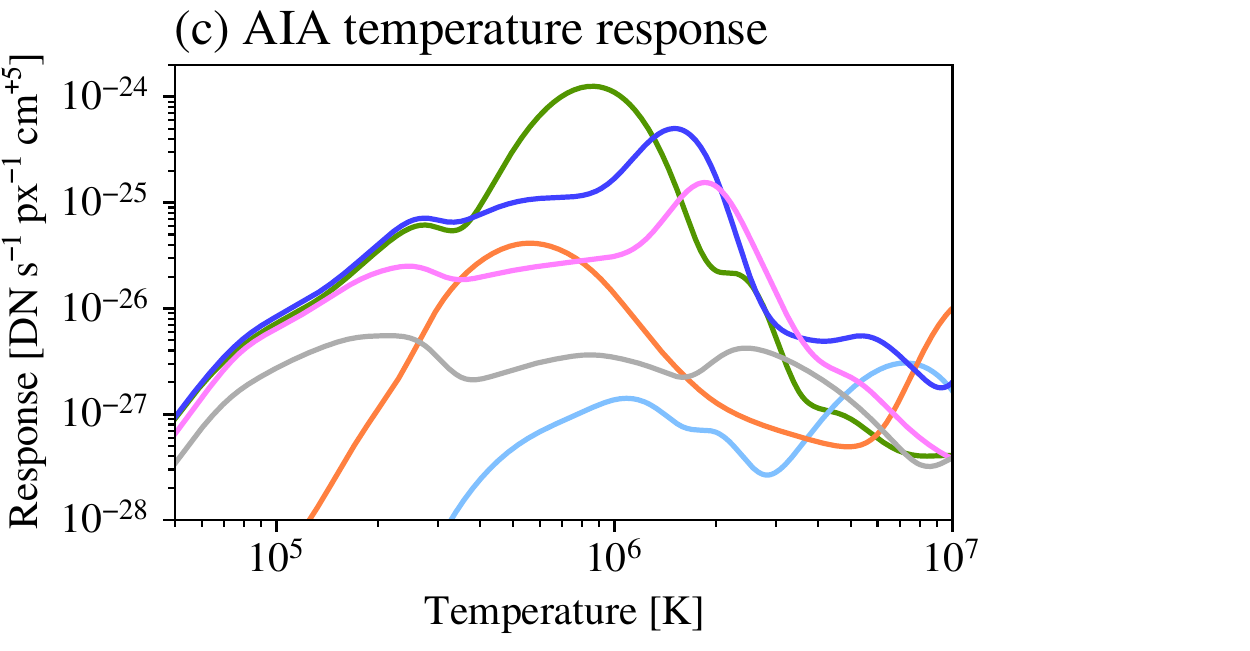}
\caption{
Synthetic intensities for the simulation that best reproduces the observations of the \rainbow.
(a)~Intensity in the
\SI{171}{\angstrom} (\textit{red}),
\SI{193}{\angstrom} (\textit{green}), and
\SI{335}{\angstrom} (\textit{blue})
bands of \SDO/AIA as a function of time and position along the loop.
The black contour shows regions where the temperature is lower than \SI{0.5}{MK}.
(b)~Intensities in the six coronal channels of AIA, integrated close to the apex, ($\SI{370}{Mm} < s < \SI{380}{Mm}$).
Each light curve is normalised by subtracting its average and dividing by its standard deviation.
(c)~Temperature-response function of the \SDO/AIA channels shown on panel~(b).
}
\label{fig:simu_rainbow_intensity}
\end{figure}

Additionally, we use the simulations to determine the heating parameters and the cross-sectional area expansion of the \rainbow.
To that end, we search for a simulation that uses the extrapolated geometry (\textit{S10}--\textit{S98}), has TNE cycles with abundant formation of coronal rain, and has the closest period to that of the observed pulsations (\SI{6.7}{h}).
The simulation which best matches these criteria has a relative area expansion of \num{60} (geometry \textit{S60}),
and a symmetric, highly stratified heating ($\lambda_1 = \lambda_2 = \SI{15}{Mm} = \num{0.024} L$)
with a high heating rate ($H_1 = \num{20480} H_0 = \SI{307}{\micro\watt\per\cubic\meter}$).
It displays TNE cycles with a period of \SI{8}{h} and abundant formation of coronal rain.
The simulations using the extrapolated geometries \textit{S10} to \textit{S60} and the aforementioned heating parameters ($\lambda_1$, $\lambda_2$ and $H_1$) all have similar behaviour and period.
However, simulations that use other heating parameters display periods of at least \SI{10}{h}.
Hence, the heating parameters are well constrained, while the loop area expansion is not.

In \autoref{fig:simu_rainbow_t_ne_v}, we show the evolution of the temperature~(a), electron number density~(b), and velocity~(c) as a function of position along the loop and time, during two pulsation periods.
The temperature and density profiles show that a large condensation periodically forms at the top of the South leg (around $s = \SI{375}{Mm}$, which is about \SI{25}{Mm} away from the South edge of the apex dip).
This condensed plasma is then evacuated along the same footpoint, thus resulting in coronal rain.
While the condensation falls, a zone of lower density and higher velocity forms at the opposite footpoint.
When the condensation reaches the chromosphere, a smaller condensation bounces back up, but does not reach the loop apex.
A perturbation travelling at the sound speed is also created, which reaches the opposite footpoint.

In \autoref{fig:simu_rainbow_t_ne_v}~(d), we show the evolution of the temperature, electron number density, and velocity at the location where the condensation forms (near the apex, for $\SI{370}{Mm} < s < \SI{380}{Mm}$), and in the leg along which it falls ($\SI{500}{Mm} < s < \SI{510}{Mm}$).
Near the apex, the cycle starts with a temperature of \SI{0.8}{MK} and a density of $\SI{e14}{m^{-3}}$.
The plasma starts to condense at the apex, with the temperature decreasing to \SI{0.1}{MK} while the density increases to $\SI{7e14}{m^{-3}}$.
The condensed plasma then starts falling along the loop (the apex velocity reaches \SI{25}{\kmps}).
As it falls along the loop, the plasma continues to condense and its velocity increases.
In the loop leg, the temperature decreases to \SI{0.05}{MK}, the density increases to $\SI{1e16}{m^{-3}}$, and the velocity increases to \SI{100}{\kmps}.

The condensation starts falling with a constant acceleration of about \SI{5}{\mpssq} along the loop.
The acceleration then increases in the loop leg (for $s > \SI{420}{Mm}$, where the loop is steeper), and reaches a maximum value of \SI{40}{\mpssq}.
\citet{AuchereEtAl2018} measured coronal rain acceleration in the \rainbow loop, and reported values ranging from \SI{5}{\mpssq} to \SI{10}{\mpssq} on the plane of the sky\footnote{The angle between the loop and the plane of the sky (POS) about \ang{10;;} in this section of the loop. Thus, the accelerations projected onto the POS is about the same as along the loop.}.
While the simulation features a larger acceleration in the lower part of the loop, it is overall consistent with the observations.
In particular, these values are about five times lower than the gravitational acceleration projected along the loop (\SI{26}{\mpssq} where the condensation starts falling, and \SI{235}{\mpssq} it reaches maximum acceleration).
The observed and simulated acceleration profiles are typical of coronal rain, which is known to have velocities lower than the free-fall speed \citep[see \eg][]{Schrijver2001, DeGroofEtAl2005, AntolinEtAl2010, VashalomidzeEtAl2015}.
This happens because coronal rain is slowed down by forces that oppose gravity, such as pressure gradient \citep{OliverEtAl2014}, or drag \citep[in multidimensional simulations, ][]{Martinez-GomezEtAl2020}.

The period of the TNE cycles in this simulation is \SI{20}{\percent} longer than the one observed in the \rainbow. However, this is the simulation using stereoscopically-reconstructed geometries with the smallest period.
\citet{JohnstonEtAl2019} have shown that the period of TNE pulsations can increase if the background heating ($H_0$ in our case) is too high.
They report a period increase for background heating values above $\SI{e3}{\nano\watt\per\cubic\meter}$.
Although we did not include different values of $H_0$ in our parametric study, the background heating values that we used ($< \SI{55}{\nano\watt\per\cubic\meter}$) should thus be low enough to not modify the TNE periods.

\begin{table*}
\footnotesize
\caption{Event type counts in all simulations of the \rainbow.
For each geometry, we give the number of simulations showing each event type (\textit{Count} column),
the fraction compared to the total number of simulations using the geometry (\textit{Frac. tot.} column),
and the fraction compared to the number of simulations showing TNE (\textit{Frac. TNE} column).
}
\label{tab:event_type_count}
\centering
\begin{tabular}{l|rrr|rrr|rrr}
\hline
\hline
                                         &  \multicolumn{3}{c|}{\textit{S10} to \textit{S98}}          &  \multicolumn{3}{c|}{\textit{E98}}                         &  \multicolumn{3}{c}{\textit{E13}}                          \\
                                         &  Count       &  Frac. tot.           &  Frac.         TNE     &  Count      &  Frac.         tot.   &  Frac.         TNE     &  Count      &  Frac.         tot.   &  Frac.         TNE    \\
\hline
TNE behaviours: \bwh                     &              &                       &                        &             &                       &                        &             &                       &                        \\
\hspace{1em} prominence \bpr             &  \num{1063}  &  \SI{15.7}{\percent}  &  \SI{20.0}{\percent}   &        --   &                   --  &                   --   &         --  &                   --  &                   --   \\
\hspace{1em} cycle with rain \bcc        &  \num{3658}  &  \SI{54.2}{\percent}  &  \SI{69.0}{\percent}   &  \num{665}  &  \SI{59.1}{\percent}  &  \SI{87.3}{\percent}   &  \num{344}  &  \SI{30.6}{\percent}  &  \SI{88.2}{\percent}   \\
\hspace{1em} cycle without rain \bic     &  \num{ 582}  &  \SI{ 8.6}{\percent}  &  \SI{11.0}{\percent}   &  \num{ 97}  &  \SI{ 8.6}{\percent}  &  \SI{12.7}{\percent}   &  \num{ 46}  &  \SI{ 4.1}{\percent}  &  \SI{11.8}{\percent}   \\
\hspace{1em} (all) \bwh                  & (\num{5303}) & (\SI{78.5}{\percent}) &                        & (\num{762}) & (\SI{67.7}{\percent}) &                        & (\num{390}) & (\SI{34.7}{\percent}) &                        \\
No TNE \bno                              &  \num{1227}  &  \SI{18.2}{\percent}  &                        &  \num{300}  &  \SI{26.7}{\percent}  &                        &  \num{735}  &  \SI{65.3}{\percent}  &                        \\
Failed simulations \bfa                  &  \num{ 220}  &  \SI{ 3.3}{\percent}  &                        &  \num{ 63}  &  \SI{ 5.6}{\percent}  &                        &         -- &                    -- &                        \\
\hline
\end{tabular}
\end{table*}

To better compare to the observations of \citet{AuchereEtAl2018}, we compute synthetic intensities for this simulation.
The intensities are computed using the \SDO/AIA temperature response functions from Chianti 8 \citep{DereEtAl1997-chianti1, DelZannaEtAl2015-chianti14v8}, and the method described in \citet[Section 3.2.1]{FromentEtAl2017}.
In \autoref{fig:simu_rainbow_intensity}~(a), we show the evolution of the intensity in the intensity in the \num{171}, \num{193}, and \SI{335}{\angstrom} bands of AIA, as a function of the position along the loop and time.
In \autoref{fig:simu_rainbow_intensity}~(b), we plot the evolution of the intensity integrated close to the apex (between \num{370} and \SI{380}{Mm}) for the six coronal channels of AIA\footnote{We note that the choice of the integration region does not significantly influence the order in which different bands peak.  % 237-397 gives the same order as 370-380 Mm.
}.
The temperature-response function of these bands is shown in \autoref{fig:simu_rainbow_intensity}~(c).
The intensity first peaks in the \SI{171}{\angstrom} band (maximum temperature response at \SI{0.9}{MK}), which is consistent with the fact that the loop reaches a maximum temperature of \SI{0.8}{MK} during the cycle.
The intensity then peaks in the other bands in the following order:
\SI{94}{\angstrom},
\SI{131}{\angstrom},
\SI{193}{\angstrom},
\SI{211}{\angstrom}, and
\SI{335}{\angstrom}.
These peaks, except \SI{94}{\angstrom}, correspond to local maximums of each band’s temperature response function below \SI{0.8}{MK}, peaking in order of decreasing temperature.
This behaviour is consistent with the numerous reports that coronal loops (undergoing TNE or not) are generally observed in their cooling phase \citep{WarrenEtAl2002, WinebargerEtAl2003, WinebargerWarren2005, Ugarte-UrraEtAl2006, Ugarte-UrraEtAl2009, Mulu-MooreEtAl2011, ViallKlimchuk2011, ViallKlimchuk2012, FromentEtAl2015}.
However, the bands peak in a different order in the observations of the \rainbow, with \SI{193}{\angstrom}, \SI{94}{\angstrom}, \SI{211}{\angstrom}, and \SI{335}{\angstrom} peaking before \SI{171}{\angstrom} \citep{AuchereEtAl2018}.
While this order is also consistent with loops observed in their cooling phase, it indicates that the maximum temperature observed in the \rainbow ranges between \num{1} and \SI{1.5}{MK}, thus higher than in our simulation.
This temperature difference would explain why the simulated AIA bands peak in a different order than in the observations.

The simulation that best matches the observed behaviour of the \rainbow reproduces its main observed characteristics: the development of TNE cycles with abundant formation of coronal rain.
While it does not precisely reproduce some quantitative details of the observation (the pulsation period and the maximum temperature), the qualitative behaviour is similar.
This allows us to conclude that a highly stratified, symmetric heating is required in order to produce the periodic coronal rain observed in the \rainbow.
We can also conclude that these loops must have a large cross-sectional area expansion.

~\\

\section{The role of asymmetries in the formation of coronal rain}
\label{sec:role_of_asymmetries}

\subsection{Simulation statistics}
\label{subsec:sim_statistics}

To get an overview of the different kind of events obtained in the simulations of the \rainbow, we count the number of simulations showing each kind of behaviour: TNE cycles with and without rain, formation of a prominence-like structure, relaxation to a steady-state (\ie no TNE), and failed simulation.
The counts are given in \autoref{tab:event_type_count}.
In this table, we also include the fraction of the explored parameter space corresponding to each type of event.
Because these values depend on the choice of the parameter space, they do not constitute a precise measurement of the expected event type.
Rather, they offer a synthetic view of the trends visible in \autoref{fig:full_study}.

We first consider the \num{6750} simulations performed with the six stereoscopically-reconstructed geometries (\textit{S10} to \textit{S98}).
These simulations undergo TNE in a large fraction of the explored parameter space (\SI{78.5}{\percent}).
They are split in three groups:
cycles with coronal rain (\SI{69.0}{\percent} of simulations with TNE, or \SI{86.3}{\percent} of simulations with TNE cycles),
formation of a prominence-like structure (\SI{20.0}{\percent}),
and cycles without rain (\SI{11.0}{\percent}).
In \SI{18.2}{\percent} of the parameter space, simulations relax to a steady state, thus showing no TNE behaviour.
Finally, \SI{3.3}{\percent} of the simulations fail because of the numerical issues discussed in \autoref{subsec:sim_numerical_setup}.

The \num{1125} simulations performed with the extrapolated geometry \textit{E98} have relatively similar statistics.
A slightly smaller fraction of the parameter space experiences TNE (\SI{67.7}{\percent}), and no simulation forms a prominence-like structure.
However, \SI{87.3}{\percent} of TNE cycles produce coronal rain, which is very close to the value obtained for geometries \textit{S10} to \textit{S98} (\SI{86.3}{\percent}).
Besides, more simulations, albeit still a small fraction, fail because of numerical issues (\SI{5.6}{\percent}).

Finally, the \num{1125} simulations using the extrapolated geometry \textit{E13} have rather different statistics.
A much smaller fraction of the parameter space experiences TNE (\SI{34.7}{\percent}).
Nonetheless, a similar fraction of cycles produce coronal rain (\SI{88.2}{\percent}).
None of the simulations using this geometry fail because of numerical issues.
These differences are explained by the fact that the length and altitude profile of geometry \textit{E13} strongly differ from those of the other geometries (see \autoref{sec:rainbow_geometry}).

Overall, a large fraction of the parameter space that we explored experiences TNE.
Periodic coronal rain is easily formed in the \rainbow, with on average \SI{86}{\percent} of TNE cycles forming coronal rain, for the parameters we explored.

\subsection{Interplay between the asymmetry of the loop and of the heating}
\label{subsec:interplay_asym_loop_heating}

We have shown that TNE cycles with coronal rain occur preferentially in the simulations of the \rainbow.
Indeed, they are the most common behaviour, found in \SI{51}{\percent} of all simulations, and \SI{86}{\percent} of simulations with TNE cycles.
The simulated behaviour is consistent with the observation of abundant periodic coronal rain showers (the “monsoon”) in the \rainbow loop \citep{AuchereEtAl2018}.

\begin{figure}
\includegraphics[width=.74\columnwidth]{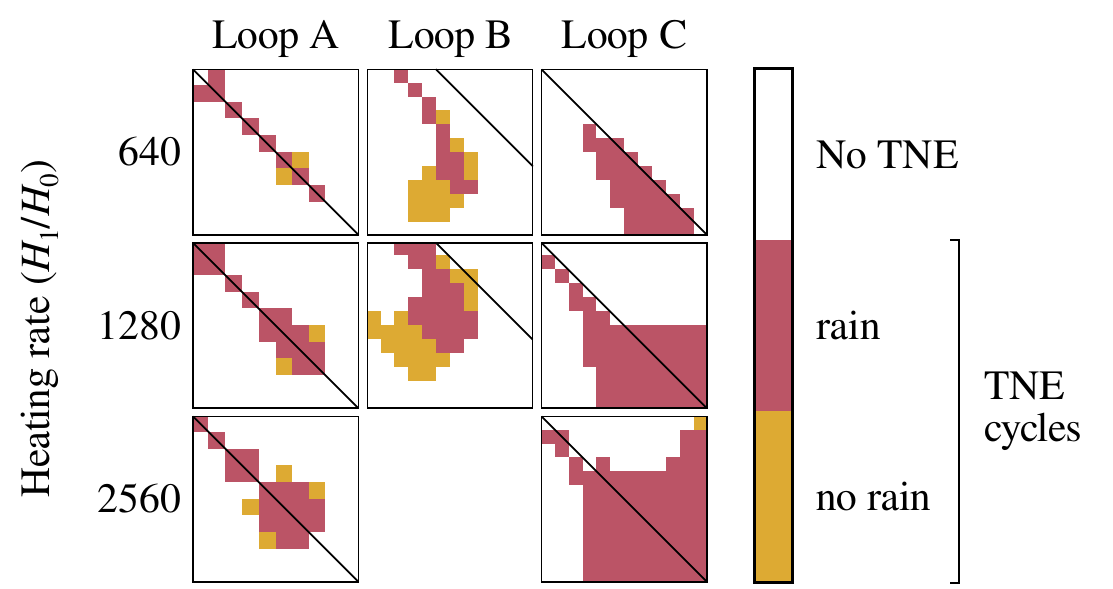}
\caption{
Event classification for the simulations performed by \citet{FromentEtAl2018}, shown with a similar layout as \autoref{fig:full_study}~(d).
\textit{Loop~A} is a semicircular with a length of \SI{367}{Mm}, heated with 10 values of $\lambda_1$ and $\lambda_2$, both ranging from \num{2} to \SI{11}{\percent} of the loop length.
\textit{Loop~B} is a highly asymmetric loop with a length of \SI{367}{Mm}, heated with 12 values of $\lambda_1$ and $\lambda_2$, ranging respectively from \num{7} to \SI{18}{\percent}, and from \num{2} to \SI{13}{\percent} of the loop length.
\textit{Loop~C} is a moderately asymmetric loop with a length of \SI{139}{Mm}, heated with 12 values of $\lambda_1$ and $\lambda_2$, both ranging from \num{4} to \SI{15}{\percent} of the loop length.
The diagonal lines in each square correspond to $\lambda_1 = \lambda_2$.
}
\label{fig:cf2018_event_type}
\end{figure}

While the \rainbow loop has characteristics similar to those of the loop studied by \citet{FromentEtAl2015}, this latter had little or no rain.
In order to understand this different behaviour, we compare the simulations of the \rainbow to the simulations performed for three different geometries by \citet{FromentEtAl2018}.
The authors used the same simulation setup and heating function as in the current paper.
They consider a semi-circular loop geometry with a length of \SI{367}{Mm} (\textit{loop~A}),
a very asymmetric loop obtained from magnetic field extrapolation (\textit{loop~B}, matching the observations for which no rain was detected),
and a relatively symmetric loop also obtained from extrapolation (\textit{loop~C}).
\textit{Loop~B} corresponds to a loop bundle in which TNE cycles with little to no coronal rain had been previously identified \citep{FromentEtAl2015}.
Its eastern leg (at $s = 0$) has a high inclination, resulting in a very small projected gravity in the first \SI{10}{\percent} of the loop length \citep[][Figure 3]{FromentEtAl2018}.
\textit{Loop~C} is a loop in which no TNE behaviour was observed.
The legs of this loop are relatively symmetric.
The classification of events \citet{FromentEtAl2018} obtained for these loops is reproduced in \autoref{fig:cf2018_event_type} for easier comparison with our simulations of the \rainbow (\autoref{fig:full_study}~d).

The simulations performed by \citet{FromentEtAl2018} for the two symmetric geometries preferentially produce coronal rain (\SI{84.6}{\percent} TNE of cycles in \textit{loop~A} and \SI{99.4}{\percent} of cycles in \textit{loop~C} produce rain).
However, less rain forms in the very asymmetric \textit{loop~B} (only \SI{50.7}{\percent} of the cycles produce rain).

This comparison between the simulations of the \rainbow and the simulations of \citet{FromentEtAl2018} suggests that the geometry of the loop has, on its own, a significant effect on the formation of coronal rain during TNE cycles.
In a relatively symmetric loop, coronal rain forms for a wider range of heating functions.
On the other hand, cycles without rain occur more easily in very asymmetric loops.\footnote{We verified that these conclusions are unchanged if we only consider the simulations by \citet{FromentEtAl2018} and of the \rainbow with overlapping heating parameters.}
However, the heating function also plays a key role on the formation of rain.
In the corona, a given loop is only subject to one of the many explored heating functions.
The development of TNE thus results from a specific match between the loop geometry and the heating function, as noted by \citet{FromentEtAl2018}.
Our simulations of the \rainbow support this conclusion, and further suggest that symmetric loops more frequently produce cycles with coronal rain, while asymmetric loops lead to more cycles without rain.
In the case of the \rainbow, cycles with no rain only occur in a small volume of the parameter space.
This makes such cycles very unlikely to occur in a real coronal loop, as it would require the heating and geometry to be stable over several days.
On the other hand, cycles with rain could be produced even with varying heating or geometry, provided that the variations are slow enough.

Overall, coronal rain is formed if the condensing plasma cools down to chromospheric temperatures before being evacuated from the loop.
In this case, the condensed plasma then falls along the loop as coronal rain.
On the contrary, no rain is formed if the condensations are destabilised earlier, and fall while still at coronal temperatures.
In relatively symmetric loops (\eg the \rainbow), condensations can remain longer at the loop apex for a wide variety of heating functions.
Only very asymmetric heating functions lead to cycles with no rain.
As a result, such loops preferentially form cycles with rain.
In asymmetric loops (\eg \textit{loop~B} of \citealp{FromentEtAl2018}), condensations can only remain at the apex if an oppositely asymmetric heating creates upflows that compensate the loop's asymmetry.
In such loops, rain is overall less likely to occur, as it can only form under specific heating conditions.

Our simulations of the \rainbow and the simulations of \citet{FromentEtAl2018} are also consistent with the qualitative predictions of the analytical formulas of \citet{KlimchukLuna2019} which provide conditions for TNE (with or without rain) to occur in coronal loops.
We compared the simulations to the overall predictions of these formulas, but did not perform a quantitative comparison.
Their first condition \citep[][equation 13]{KlimchukLuna2019} is that the heating must be concentrated at low altitudes. 
This is the case in all simulations.
The second condition \citep[][equation 34]{KlimchukLuna2019} is that the product of the asymmetries in the heating and the cross-sectional area must not be too important.
In the case of the \rainbow, the loop shape and area expansion are relatively symmetric.
As a result, a wider variety of heating functions, including highly asymmetric ones, may lead to TNE.
On the other hand, \textit{loop~B} of \citet{FromentEtAl2018} has a very asymmetric shape and area expansion.
This imposes a larger constraint on the heating function: in order to have TNE, the heating must also be asymmetric in order to compensate the geometric asymmetry of the loop.

\subsection{Other parameters that may influence the formation of coronal rain}
\label{subsec:other_parameters}

In the current paper, we explore different loop geometries and heating parameters, and their influence on the development of TNE in coronal loops.
However, other factors may influence the evolution of plasma in a loop.
In particular, we did not consider any variation of the heating or the geometry over time.
Variations of the heating over short time scales (\ie less than the loop cooling time) have been considered in a few studies of TNE.
\citet{AntolinEtAl2010} and \citet{SusinoEtAl2010} have shown that complete plasma condensations can occur in impulsively heated loops.
\citet{WinebargerEtAl2018} and \citet{JohnstonEtAl2019} have studied the influence of impulsive heating on some characteristics of TNE cycles.
Most of these studies consider heating functions with a constant waiting time between the pulses.
It would also be worth considering stochastic heating functions, as these better describe nanoflare heating mechanisms in the solar corona.
Stochastic heating functions have been considered in several simulations of coronal loops (\eg, in 0D by \citealp{Cargill2014, CargillEtAl2015, BarnesEtAl2016b}, and in 1D by \citealp{CargillEtAl2015, BradshawViall2016}), but only in a few simulations of TNE cycles \citep{AntolinEtAl2010, Antolin2020}.
Including such impulsive stochastic heating functions in large scale parametric studies is an essential next step to understand the relationship between heating and TNE cycles.

As noted in previous studies of TNE cycles \citep{AntolinEtAl2010, AuchereEtAl2018, FromentEtAl2018}, the heating parameters and loop geometry are also likely to vary over longer time scales (\ie longer than the loop cooling time), thus resulting in a variation of the TNE conditions over time.
This could result in non-periodic TNE cycles, cycles only lasting for a few periods, or intermittent production of coronal rain from one cycle to another.
Understanding the range of behaviours that can stem from such long-term variations would require the exploration of a much larger parameter space than considered in the current study.
In the case of the \rainbow, one could take into account the evolution of the loop geometry over time.
Doing so would require the reconstruction of the loop geometry at several time intervals, which would be an even greater challenge than the reconstruction presented in \autoref{sec:rainbow_geometry}.

The conclusions presented in this paper are based on simulations of realistic loop geometries, which were derived from observations.
The symmetric and asymmetric loops result in clearly distinct TNE evolutions.
However, it is unclear whether the formation of coronal rain is constrained by asymmetries of specific geometric features (such as the cross-sectional area, or the slope of the loop legs), or results from any geometric asymmetry.
We note that the asymmetric \textit{loop~B} of \citet{FromentEtAl2018} has a unique geometry, with an asymmetric cross-sectional area and one very inclined leg
This leg is almost horizontal over its first \SI{30}{Mm}, and partly lays in the chromosphere.
We also note that the loops considered by \citet{FromentEtAl2018} and in the current paper have very different area expansion profiles (\autoref{fig:rainbow_2d}~c).
However, this is unlikely to have a significant influence compared to the loop asymmetry, because \textit{loops}~\textit{A} and \textit{C} also preferentially produce coronal rain, despite having different expansion profiles than the \rainbow loops.
In order to understand whether the formation of coronal rain is influenced by asymmetries in specific parts of the loop, it would be interesting to realise simulations for a wider variety of loop geometries. 
The use of synthetic geometries would allow to better control the different geometric features.

\section{Summary}
\label{sec:summary}

In this paper, we present 1D hydrodynamic simulations of the “\rainbow”, a periodic coronal rain event observed by \citet{AuchereEtAl2018}.
This event produces abundant coronal rain showers, which strongly contrasts with previous observations of TNE cycles that did not show evidence of coronal rain \citep{FromentEtAl2015}.

To understand which parameters influence the formation of coronal rain during TNE cycles, we performed 9000 simulations of the \rainbow, exploring different heating parameters (volumetric heating rate and asymmetry), and variants of the geometry (shape and area expansion of the loop).
To perform these simulations, we reconstructed the three-dimension geometry of the loop.

These simulations display the three behaviours that can result from TNE in coronal loops: evaporation and condensation cycles with coronal rain, cycles without coronal rain, and formation of a prominence-like structure.
We thus show that prominences and TNE cycles can develop in similar conditions (same geometry, and similar heating parameters).
This constitutes a new result for the understanding of TNE in the solar corona.

By searching for the simulation that best reproduces the observed characteristics of the \rainbow, we estimate the heating conditions of this event.
The \rainbow event results from an intense and symmetric heating localised near the footpoints of the loop (heating scale-height of \SI{15}{Mm} in both legs).
Furthermore, a small variation of the heating function results in the formation of a prominence-like structure.
This would also be consistent with the onset prominence formation reported by \citet{AuchereEtAl2018}.

Finally, we use the large parametric study to explore the relationship between the geometry of the loop, the asymmetry of the heating, and the formation of coronal rain during TNE cycles.
We compare the simulations of the \rainbow (symmetric loop) with another set of simulations performed by \citet{FromentEtAl2018} (two symmetric loops, and one asymmetric loop).
We conclude that coronal rain is overall more likely to occur in relatively symmetric loops.
In such loops, rain forms for a wide range of heating functions.
In asymmetric loops, however, rain can only form if the heating precisely compensates the geometric asymmetry, so that condensations can remain in the corona long enough to reach chromospheric temperatures.
When the heating does not compensate the loop asymmetry, condensations are evacuated from the loop before they can cool down to chromospheric temperatures, resulting in cycles with no coronal rain.

\citet{FromentEtAl2018} already noted that TNE requires a specific match between the loop geometry and the heating conditions.
The simulations of the \rainbow suggest that a symmetric loop allows for a looser match between the geometry and the heating.
Both behaviours are consistent with the qualitative predictions of the formulas of \citet{KlimchukLuna2019}.

\begin{acknowledgements}
We thank Peter Cargill for his in depth referee review, which significantly improved the manuscript.
G.P. has received funding from the European Research Council (ERC) under the European Union’s Horizon 2020 research and innovation program (grant agreement No. 724326).
We acknowledge support from the International Space Science Institute (ISSI), Bern, Switzerland to the International Team 401 “Observed Multi-Scale Variability of Coronal Loops as a Probe of Coronal Heating”.
AIA and HMI data are courtesy of NASA/SDO and the AIA and HMI science teams.
EUVI data are courtesy of the STEREO SECCHI team.
This work used data provided by the MEDOC data and operations centre (CNES/CNRS/Univ. Paris-Sud), \href{http://medoc.ias.u-psud.fr}{medoc.ias.u-psud.fr}.
CHIANTI is a collaborative project involving George Mason University, the University of Michigan (USA), University of Cambridge (UK) and NASA Goddard Space Flight Center (USA).
\textit{Software:}
Astropy \citep{Astropy2013, Astropy2018},
ChiantiPy \citep{Dere2013-chiantipy},
JHelioviewer \citep{MullerEtAl2017},
SolarSoft \citep{FreelandHandy2012}.
\end{acknowledgements}

\appendix

\section{Stereoscopic reconstruction}
\label{appendix:geometry_stereo}

We reconstructed the shape of the \rainbow loop \citep{AuchereEtAl2018} using stereoscopic observations of \SDO/AIA and \STEREO/SECCHI/EUVI.
In this appendix, we summarise the method and the results pertinent to the present work.
More details will be given in a forthcoming companion paper.

The \rainbow event was observed at the limb by AIA during \SI{2.5}{days} starting from \DTMdate{2012-07-23}, \DTMtime{00:00:00}.
During this period, the \STEREOB probe had a separation of \ang{115;;} with Earth, and observed the event on-disk.
The event was not visible from \STEREOA.
The EUVI data consist of images in the \SI{171}{\angstrom} band with a \SI{120}{min} cadence, and in the \SI{304}{\angstrom} band with a cadence of \SI{2.5}{min} for the first \SI{200}{min} and \SI{10}{min} afterwards.
The temperature response function of the \SI{171}{\angstrom} band peaks at \SI{0.9}{MK}, which allows the observation of plasma at coronal temperature.
The \SI{304}{\angstrom} band peaks at \SI{0.09}{MK}, allowing the observation of coronal rain.
During the same time period, AIA images are available in these two bands with a cadence of \SI{12}{s}.

The coronal rain is easily observed at the limb in the AIA \SI{304}{\angstrom} images.
We use a combination of intensity thresholding and manual clean-up to identify the regions of the images which contain coronal rain.
The coronal rain is harder to see on-disk in the EUVI \SI{304}{\angstrom} images.
To better detect it, we divide the intensity of each image by the average of the 20 previous and 20 following images.
The resulting intensity maps are then used to manually select the regions where coronal rain is observed.

For each pair of \SI{304}{\angstrom} images from \SDO and \STEREOB, we thus obtain two binary masks, containing the regions in which coronal rain is observed.
In order to reconstruct the longest possible fraction of the loop, we select a pair of masks in which the rain occupies a large fraction of the loop seen in the \SI{171}{\angstrom} images.
The selected images were captured on \DTMdate{2012-07-25}, \DTMtime{03:06:47}, near the end of the observing sequence.
We then use the two masks to reconstruct the volume occupied by the coronal rain.
To that end, we trace the lines of sight of each rain pixel in the masks, and compute the intersection between the lines of sight from \SDO and from \STEREOB.
Such a stereoscopic reconstruction technique assumes that both instruments observe the same opaque object.
However, that is not entirely the case here: only the dense core of the rain is visible on-disk in absorption from \STEREOB, while a larger volume of the rain is seen off-limb in emission from \SDO.
Additionally, ambiguities occur when rain is detected in adjacent loop threads.
When both instruments observe rain in two adjacent threads, four distinct threads are present in the volume reconstruction.
But only two of these threads are real.
We resolve such ambiguities manually, by ensuring the continuity of threads along the loop, and in time.
The loop footpoints are inferred manually, as they are partially hidden behind the limb in the AIA image, and are difficult to see in the EUVI images in which the rain forms very thin strands.
To that end, we extend the volume occupied by the coronal rain down to the solar surface, while trying to preserve the overall shape of the loop.
This affects the lowest \SI{5}{Mm} of the northern leg, and the lowest \SI{20}{Mm} of the southern leg.
Finally, we fit a spline to the remaining voxels.
This yields an approximation of the shape of the \rainbow loop.
The resulting 3D shape is shown in \autoref{fig:rainbow_3d}.
The altitude and projected gravity and are shown in \autoref{fig:rainbow_2d}~(a) and (b).

The stereoscopic reconstruction reveals a dip near the apex of the loop.
While this feature is unusual for coronal loops, it is consistent with some prominence models \citep[\eg][]{PriestEtAl1989}.
Such dip is consistent with the observations of \citet{AuchereEtAl2018}, who report that the condensations linger at the apex before falling, and suggest a possible link with prominence formation.
This dip appears to be a robust result from the reconstruction, and we are rather confident that it is not an artefact.
The dip is located at the centre of the loop ($s = \SI{319}{Mm}$), and at the altitude $z = \SI{118}{Mm}$.
The southern edge of the dip is located very close to the dip, both in position along the loop (\SI{32}{Mm} away, at $s = \SI{351}{Mm}$), and in altitude (\SI{2}{Mm} higher, at $z = \SI{120}{Mm}$).
The northern edge is further from the dip along the loop (\SI{87}{Mm} away, at $s = \SI{232}{Mm}$) and in altitude (\SI{13}{Mm} higher, at $z = \SI{131}{Mm}$).
On this northern side, the altitude reaches \SI{120}{Mm} (the height of the South edge)at $s = \SI{298}{Mm}$.
Overall, the dip has a depth of \SI{2}{Mm} and a width of \SI{53}{Mm}.

\section{Magnetic field extrapolation}
\label{appendix:geometry_extrapolation}

\begin{figure}
\includegraphics[width=\columnwidth]{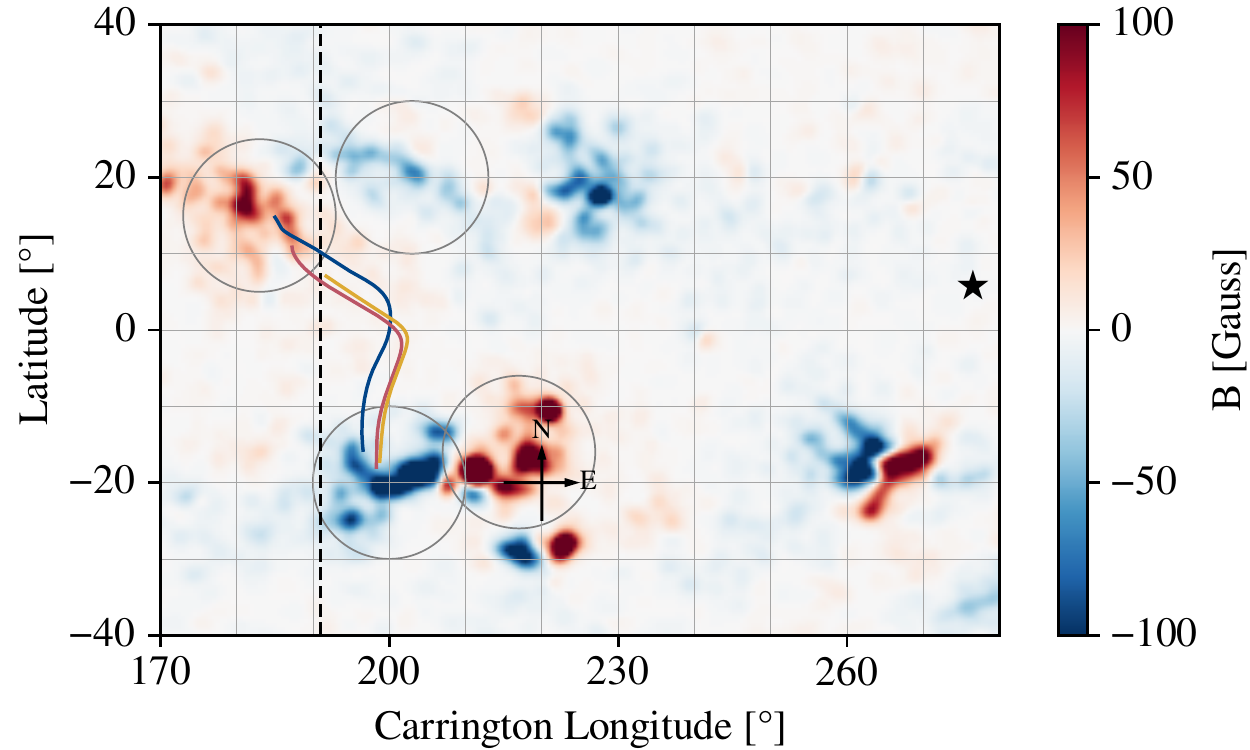}
\caption{
  Magnetogram used for the magnetic field extrapolation.
  The vertical dashed line at the longitude of \ang{191;;} corresponds to the limit between the synoptic magnetograms from rotations CR2125 (left) and CR2126 (right).
  The quadrupolar structure mentioned in the text is marked by grey circles.
  The black star indicates the solar disk centre as seen from Earth at the time for which geometry \textit{S} was reconstructed (\DTMdate{2012-07-25}, \DTMtime{03:06:47}).
  The loop geometries, the heliographic grid, and direction cross shown in \autoref{fig:rainbow_3d} are reproduced for easier comparison.
  }
\label{fig:rainbow_magnetogram}
\end{figure}

\begin{table}
\footnotesize
\caption{
Properties of the extrapolated field lines used to generate area expansion profiles for the stereoscopic loop shape \textit{S}.
Length $L$, magnetic field at the North footpoint $B(0)$, at the South footpoint $B(L)$, and minimum value close to the apex $B_\mathrm{min}$ .
}
\label{tab:mag_lines_properties}
\begin{tabular}{lcccc}
\hline
\hline
  \shortstack{Line} &
  \shortstack{$L$ [Mm]} &
  \shortstack{$B(0)$ [G]} &
  \shortstack{$B_\mathrm{min}$ [G]} &
  \shortstack{$B(L)$ [G]} \\
\hline
\textit{E08} & {430.63} &  {5.06} & {0.60} & {44.78} \\
\textit{E13} & {465.54} &  {8.59} & {0.65} & {47.02} \\
\textit{E25} & {492.24} & {23.45} & {0.95} & {48.37} \\
\textit{E98} & {628.86} & {41.92} & {0.43} & {62.68} \\
\hline
\end{tabular}
\end{table}

To determine the area expansion of the \rainbow loop \citep{AuchereEtAl2018}, we perform a potential magnetic field extrapolation from a synoptic line of sight magnetogram of \SDO/HMI.

The synoptic magnetogram has a resolution of \SI{0.5}{\degree\per\pixel} in Carrington coordinates. 
The magnetic field at each longitude was measured when the corresponding meridian was as the centre of the disk as seen from \SDO (\ie approximately from Earth).
For the Carrington longitudes between \ang{0;;} and \ang{191;;}, we use magnetograms recorded at the rotation before the observation of the \rainbow (CR2125), that is about 20 days before.
For longitudes between \ang{191;;} and \ang{360;;}, we use magnetograms from the following rotation (CR2126), recorded a few days after the observation of the \rainbow.
The boundary between the magnetograms from both rotations is thus located between the footpoints of the stereoscopically-reconstructed shape \textit{S} (at Carrington longitudes \ang{185;;} and \ang{207;;}, respectively).
The resulting magnetogram is shown in \autoref{fig:rainbow_magnetogram}.
This composite magnetogram contains the polarities required to obtain extrapolated lines with a shape resembling that of the \rainbow: a magnetic quadrupole at the East of the loops (marked by grey circles in \autoref{fig:rainbow_magnetogram}), which explains their inclination.
This quadrupole is not present when only using data from rotation CR2125 (because the South-West negative polarity had not emerged yet), nor when only using data from rotation CR2126 (because the North-West positive polarity has decayed too much).

We then perform a potential extrapolation, where it is assumed that the magnetic field satisfies the condition $\Nabla\times\vec{B} = \vec{0}$ \citep[see, \eg][]{Regnier2013, WiegelmannEtAl2014}.
We select four magnetic field lines extrapolated from this magnetogram: \textit{E08}, \textit{E13}, \textit{E25}, and \textit{E98}.
The properties of these extrapolated lines are summarised in \autoref{tab:mag_lines_properties}.
Lines \textit{E13} and \textit{E98} are shown on Figs. \ref{fig:rainbow_3d}, \ref{fig:rainbow_2d}, and \ref{fig:rainbow_magnetogram}.
Similarly to the stereoscopically reconstructed loop shape (\autoref{appendix:geometry_stereo}), the extrapolated lines are non-planar and inclined towards the East of the solar sphere.

The footpoint magnetic field of these loops ranges from \num{5} to \SI{60}{G}, which is somewhat smaller than typical active region values (see \eg \citealp{AschwandenEtAl1999stereo_ar_loops} who report a range of \num{20}--\SI{230}{G} for a selection of 30 active region loops).
This can be explained by the low resolution of the magnetogram, which reduces and spreads out the magnetic field intensity.
Line \textit{E98} has an area expansion of \num{98}, which is higher than typical active region values (\citealp{DudikEtAl2014} report a maximum value of \num{80} in a given active region).
Overall, potential magnetic field extrapolation from a low resolution magnetogram only gives a very coarse estimation of the magnetic field, especially in active regions.
However, the main goal of these extrapolations is to estimate the relative loop cross-sectional area.
(A more precise estimation of the loop shape is already obtained from the stereoscopic reconstruction.)
Nonetheless, because of the uncertainty of this extrapolation method, we explore different cross-sectional area profiles.

We thus generate six relative loop area profiles from the extrapolated lines (\textit{E08}, \textit{E13}, \textit{E25}, and \textit{E98}).
We combine these six area profiles with the shape from the stereoscopic reconstruction \textit{S}.
The resulting geometries share the same shape, but have different area profiles, with maximum values ranging from \num{10} to \num{98.4}.
These new geometries are labelled with their maximum relative area expansion: \textit{S10}, \textit{S25}, \textit{S47}, \textit{S60}, \textit{S73}, and \textit{S98}.
Because the lines have different lengths, we scale each area expansion profile to the loop length.
The new area expansions profiles are generated from the extrapolated area expansion profiles using the following relations:
\begin{equation}
\begin{split}
& \Aline{S10} = \left[  \Aline{E08} +   \Aline{E13}\right] / 2 \\
& \Aline{S25} =         \Aline{E25} \\
& \Aline{S47} = \left[2 \Aline{E25} +   \Aline{E98}\right] / 3 \\
& \Aline{S60} = \left[  \Aline{E25} +   \Aline{E98}\right] / 2 \\
& \Aline{S73} = \left[  \Aline{E25} + 2 \Aline{E98}\right] / 3 \\
& \Aline{S98} =                    \Aline{E98}
\end{split}
\end{equation}
These cross-sectional area profiles are shown in figure \autoref{fig:rainbow_2d}~(c).

\end{document}